\documentclass[reprint,superscriptaddress,amsmath,amssymb,aps,pre]{revtex4-1}

\usepackage{graphicx}
\usepackage{dcolumn}
\usepackage{bm}
\usepackage{color}
\usepackage[makeroom]{cancel}

\usepackage{bbold}
\usepackage{soul}
\usepackage{epsfig,epic}
\usepackage{nicefrac}
\usepackage{dsfont}
\usepackage{placeins}     

\newcommand{\newc}{\newcommand}
\newc{\beq}{\begin{equation}}
\newc{\eeq}{\end{equation}}
\newc{\kt}{\rangle}
\newc{\br}{\langle}
\newc{\tr}{\text{tr}}
\newc{\eps}{\epsilon}
\newc{\beqa}{\begin{eqnarray}}
\newc{\eeqa}{\end{eqnarray}}
\newc{\longra}{\longrightarrow}
\newc{\red}{\textcolor{red}}
\newc{\blue}{\textcolor{blue}}
\newcommand{\pA}{p_A}
\newcommand{\pB}{p_B}
\newcommand{\qA}{q_A}
\newcommand{\qB}{q_B}
\newcommand{\VA}{V_A}
\newcommand{\VB}{V_B}
\newcommand{\VAB}{V_{AB}}
\newcommand{\KA}{K_A}
\newcommand{\KB}{K_B}

\providecommand*{\ui}{\text{i}}

\usepackage{color}

\newcommand{\CNtxt}{closer neighbor}
\newcommand{\CN}{\text{CN}}

\newcommand{\Marcenko}{Mar{\v c}enko}

\providecommand*{\Trace}{\text{Tr}}

\newcommand{\rhoA}{\rho^A}
\newcommand{\rhoB}{\rho^B}

\renewcommand{\Re}{\text{Re}\,}

\newcommand{\cU}{{\cal U}}

\newcommand{\HIDDEN}[1]{}

\makeatletter
\let\Hy@backout\@gobble
\makeatother

\begin{document}

\title{Eigenstate entanglement between quantum chaotic subsystems:\\
 universal transitions and power laws in the entanglement spectrum}

\author{Steven Tomsovic}
\affiliation{Max-Planck-Institut f\"ur Physik komplexer Systeme,
             N\"othnitzer Stra\ss{}e 38, 01187 Dresden, Germany}
\affiliation{Technische Universit\"at Dresden, Institut f\"ur Theoretische
             Physik and Center for Dynamics, 01062 Dresden, Germany}
\affiliation{Department of Physics and Astronomy, Washington State University,
             Pullman, WA~99164-2814}
\author{Arul Lakshminarayan}
\affiliation{Max-Planck-Institut f\"ur Physik komplexer Systeme,
             N\"othnitzer Stra\ss{}e 38, 01187 Dresden, Germany}
\affiliation{Department of Physics, Indian Institute of Technology Madras,
             Chennai, India~600036}
\author{Shashi C. L. Srivastava}
\affiliation{Max-Planck-Institut f\"ur Physik komplexer Systeme,
             N\"othnitzer Stra\ss{}e 38, 01187 Dresden, Germany}
\affiliation{Variable Energy Cyclotron Centre, 1/AF Bidhannagar,
             Kolkata 700064, India.}
\affiliation{Homi Bhabha National Institute, Training School Complex,
		 Anushaktinagar, Mumbai - 400085, India}
\author{Arnd B\"acker}
\affiliation{Max-Planck-Institut f\"ur Physik komplexer Systeme,
             N\"othnitzer Stra\ss{}e 38, 01187 Dresden, Germany}
\affiliation{Technische Universit\"at Dresden, Institut f\"ur Theoretische
             Physik and Center for Dynamics, 01062 Dresden, Germany}

\date{\today}

\begin{abstract}
\noindent

We derive universal entanglement entropy and Schmidt eigenvalue behaviors for
the eigenstates of two quantum chaotic systems coupled with a weak interaction.
The progression from a lack of entanglement in the noninteracting limit to the
entanglement expected of fully randomized states in the opposite limit is
governed by the single scaling transition parameter, $\Lambda$.  The behaviors
apply equally well to few- and many-body systems, e.g.\
interacting particles in quantum dots, spin chains,
coupled quantum maps, and Floquet systems as long as
their subsystems are quantum chaotic, and not localized in some manner.  To
calculate the generalized moments of the Schmidt eigenvalues in the
perturbative regime, a regularized theory is applied, whose leading order
behaviors depend on $\sqrt{\Lambda}$.  The marginal case of the $1/2$ moment,
which is related to the distance to closest maximally entangled state, is an
exception having a $\sqrt{\Lambda}\ln \Lambda$ leading order and a logarithmic
dependence on subsystem size.  A recursive embedding of the regularized
perturbation theory gives a simple exponential behavior for the von Neumann
entropy and the Havrda-Charv{\' a}t-Tsallis entropies for increasing
interaction strength, demonstrating a universal transition to nearly maximal
entanglement.
Moreover, the full probability densities of the Schmidt eigenvalues,
i.e.\ the entanglement spectrum, show a transition from power laws and L\'evy
distribution in the weakly interacting regime to random matrix results for the
strongly interacting regime.  The predicted behaviors are tested on a pair of
weakly interacting kicked rotors, which follow the universal behaviors
extremely well.

\end{abstract}

\maketitle

\newpage

\section{Introduction}

The entanglement properties of eigenstates of weakly interacting, but strongly
chaotic subsystems are of great interest in many different situations.  They
have been linked, for example, to emergent classical
behavior~\cite{Zur1991wnote, Zur2003}, time rate of production of entanglement
in initially separable states~\cite{Alb1992, MilSar1999a, FujMiyTan2003,
  BanLak2004, GamPat2007, TraMadDeu2008} or thermalization~\cite{Deu1991,
  Sre1994, NeiEtAl2016, AleKafPolRig2016}.  For isolated many-body systems,
thermalization manifests itself by the redistribution of initial quantum
correlations encoded in subsystems to the whole system in such a manner that it
cannot be retrieved by any experiment.  This process, called scrambling of
information, is exponential or polynomial in time, depending on whether or not
the system is chaotic.  Quantitatively this is captured for example by
out-of-time-order correlators, which measure the development of a
non-commutativity of initially commuting operators under small
perturbations~\cite{LarOvc1969, HarMal2013, LasStaHasOsbHay2013,
  SheSta2014,MalSheSta2016}.  This has been investigated theoretically and
experimentally to understand the information propagation and growth of various
entanglement measures for quantum integrable as well as quantum chaotic
many-body systems~\cite{LuiBar2017, LiFanWanYeZenZhaPenDu2017}.

All these topics are rather naturally cast into a quantum chaos framework
with which a
number of other phenomena have long been associated, such as spectral
statistics~\cite{BroFloFreMelPanWon1981,BohGiaSch1984}, e.g.\ level repulsion
and spectral rigidity, universal conductance
fluctuations~\cite{LeeSto1985,Alt1985}, eigenstate morphology being similar to
random waves~\cite{Ber1977b}, chaos-assisted
tunneling~\cite{BohTomUll1993,TomUll1994}, and quantum
ergodicity~\cite{Shn1974,CdV1985,Zel1987,ZelZwo1996,BaeSchSti1998}.

A bedrock of quantum chaotic phenomena is universality, which for our purposes
means that, with the exception of a system's fundamental
symmetries~\cite{Por1965}, essentially no information about the system is
contained in appropriately scaled local quantum fluctuation properties.  For
example, after scaling out the mean level spacing, spectral fluctuation
properties of a {\it sufficiently chaotic} system do not depend on the nature
of the system in any way, e.g.\ they are universal, and in particular,
independent of whether it is a one-body or a many-body system.  The derivation
of universal laws is quite often done with the aid of random matrix ensembles.

There are some well known exceptions to universality.  Perhaps the most
important example is localization in extended systems, whether it takes the
form of Anderson localization~\cite{And1958} or many-body localization in Fock
space~\cite{BasAleAlt2006, NanHus2015, AbaPap2017}.
This leads to an additional motivation for
understanding the universal behaviors as any deviation from universality
indicates the presence of interesting physics, such as some form of
localization or other non-ergodic phenomenon.

The concept of universality can be generalized further to incorporate the
possibility of weakly broken symmetries, which provides a powerful analysis
tool for a wide variety of problems.  In the case of breaking time reversal
invariance there is a universal
transition from the statistics of invariant systems to those with completely
broken symmetry if it is characterized as a function of the unitless transition
parameter, $\Lambda$ \cite{PanMeh1983}.
The applicability of $\Lambda$ to any symmetry,
fundamental or dynamical, describing transitions in fluctuation properties was
emphasized in Refs.~\cite{ FreKotPanTom1988,FreKotPanTom1988b}.
It is defined as the local mean square symmetry violating matrix element
divided by the mean level spacing squared, and its relevance can be deduced
from perturbation theory.  The transition parameter $\Lambda$ falls within the
interval $[0,\infty]$ with the limits being preserved symmetry and completely
broken symmetry, respectively.  The relationship of $\Lambda$ to the symmetry
breaking interaction strength depends on the details of the system under
consideration, such as the strength of interaction and the density of states,
but once the transition in some fluctuation measure is expressed as a function
of $\Lambda$, it is system independent.

Eigenstate entanglement of weakly interacting bipartite systems fits perfectly
into the generalized universality class of a dynamical symmetry breaking
nature.  Consider two sufficiently chaotic subsystems with a tunable
interaction strength.  If the interaction strength vanishes for an autonomous
Hamiltonian system there would be two constants of the motion, the energies of
each subsystem, and the dynamics of each system would be completely
independent.  A non-zero interaction strength breaks that dynamical symmetry.
Clearly, without interaction the eigenstates are product states of the
eigenstates of the subsystems, and thus completely unentangled.  In the other
extreme of a strong interaction strength, the eigenstates behave like random
states in the Hilbert space of the full system, which fluctuate about (nearly)
maximal entanglement.  The universal transition between the two extremes must
be governed by $\Lambda$ and take on a unique universal
functional form, independent of any other system properties.

The production of eigenstate entanglement between chaotic subsystems turns out
to be extremely sensitive to the interaction strength, as with all weakly
broken symmetries.  With increasing system size or complexity, less and less
interaction strength is necessary to produce eigenstates that are nearly
maximally entangled.  They are becoming statistically close to random states on
the bipartite space.  It is known that the entanglement in random states can be
used in various protocols of quantum information, including cryptography and
super dense coding~\cite{HayLeuShoWin2004,HayLeuWin2006}.  Thus, it may be
preferable to have local resources producing non-integrability and local random
states than to have local interactions that lead to near-integrable dynamics as
the later would require relatively larger non-local interaction to produce
nearly similar entanglement.

Some very useful entanglement measures are provided by the von Neumann and
Havrda-Charv\'at-Tsallis
entropies~\cite{BenBerPopSch1996,HavCha1967,Tsa1988,BenZyc2006},
denoted below as $S_{\alpha}$.
All these can be expressed as functions of the moments of
the Schmidt eigenvalues $\{\lambda_j \}$ of the reduced density matrix,
obtained after partially tracing one of the subsystems.
These eigenvalues (or their negative logarithms) have been referred to as the
``entanglement spectrum", and it has been proposed that the few most
significant of these has information about topological order in quantum Hall
states \cite{LiHal2008}. Also fluctuation properties, such as their nearest
neighbour spacing distribution, have been used to characterize complexity of
states \cite{ChaHamMuc2014}. In this paper we investigate the moments and
densities of the entanglement spectra across a complete transition, from
unentangled through perturbative regimes to that of strong coupling. For
purposes of clarity we will continue referring to the $\{\lambda_i \}$ simply
as Schmidt eigenvalues.

In Ref.~\cite{LakSriKetBaeTom2016} the universal behavior of the
first and higher order moments ($\alpha^{\text{th}}$-order) of the Schmidt
eigenvalues has been calculated, and hence all of these entropies,
as a function of $\Lambda$ using a recursively embedded and
regularized perturbation theory. The end result was
valid for the entire range of $0\le \Lambda\le \infty$, not just the
perturbation regime.  The complete derivations are given in this paper,
including new higher order contributions.

In addition, we study the limits to which the formalism can be extended, and in
particular, deal with the $\alpha=1/2$ moment.  This moment is of particular
interest as it is monotonic with the distance of the bipartite pure state
  to the closest maximally entangled one, and it is at the boundary between
moments that depend on subsystem size and those that do not.  In the quantum
information context, the ``singlet fraction"~\cite{HorHorHor1999} essentially
measures the same quantity. The statistical properties of the
Schmidt eigenvalues in the perturbative regime is also extensively
studied below and reveals the existence of power-laws and stable distributions.

Interestingly the probability that the second
largest Schmidt eigenvalue is close to the maximum possible value of $1/2$ is
non-vanishing, which implies a large number of cases where there are two
significant eigenvalues of the reduced density matrix even in the perturbative
coupling regime. A curious result is that there is a
universal function of the second largest
Schmidt eigenvalue in terms of $\Lambda$, which is suggested naturally from
perturbation theory, and has power-law tails falling as an inverse cubic.
However, the same function, but now of the difference of the largest
Schmidt eigenvalue from unity, displays the stable L\'evy distribution, having
its origin in a generalized central limit theorem, being a sum over many
heavy--tailed random variables.  The stable L\'evy distribution
is also seen to occur in the distribution of the linear entropy measure
of entanglement.

Numerical results show how this heavy--tailed density as well as the
stable L\'evy distributions are modified as the perturbation increases. In the
regime of large $\Lambda$ the largest Schmidt eigenvalue comes from a
Tracy-Widom distribution. Similar transitions are observed in the density of
eigenvalues of the reduced density matrix as it approaches the \Marcenko-Pastur
distribution for large coupling strengths.

\section{Entanglement in bipartite systems}

\subsection{Bipartite systems}

Consider a bipartite system defined on an $N_AN_B$--dimensional tensor product
space, ${\cal H}^A \otimes {\cal H}^B$, where each subsystem is defined on
$N_{A}$ and $N_{B}$--dimensional Hilbert spaces ${\cal H}^A$ and ${\cal H}^B$,
respectively.  Assume that the space is symmetry reduced, thus there are no
systematic degeneracies, and $N_{A}\le N_{B}$.  A generic situation is
described by a Hamiltonian of the form
\beq
\label{eq:Hab}
H(\epsilon)= H_A \otimes \mathds{1}_B + \mathds{1}_A \otimes H_B
        + \epsilon V_{AB},
\eeq
where $\mathds{1}_A$ and $\mathds{1}_B$ are identity operators on ${\cal H}^A$
and ${\cal H}^B$, respectively.  Alternatively, one may consider unitary
operators of the form
\beq
\label{eq:Uab}
 \cU(\eps) = (U_A \otimes U_B)\, U_{AB}(\eps),
\eeq
where $U_{AB}(\eps)\rightarrow \mathds{1}$ as $\eps \rightarrow 0$. It is
assumed that for $\eps \ne 0$ both $V_{AB}$ and $U_{AB}(\eps)$ break the
dynamical symmetry, (see \cite{ZanZalFao2000,PalLak2018:p}
for a discussion of operator
entanglement) and hence provide a genuine interaction between the two
subystems. If $\epsilon=0$, the eigenstates of $H(0)$ (or $\cU(0)$) are product
states, which are unentangled by definition.  When increasing $\epsilon>0$
the subsystems become coupled and the eigenstates become entangled.  This
transition is governed by a universal transition parameter $\Lambda$.  The
first goal is to derive the $\Lambda$-dependence of bipartite entanglement
measures for the eigenstates of $H(\eps)$ or $\cU(\eps)$, and obtain the
relationship between $\Lambda$ and $\epsilon$ within random matrix theory.

\subsection{Moments and Entropies}

To make this paper self-contained and to fix notation, some standard
definitions of the central quantities used in the following are given, see
e.g.\ \cite{NieChu2010, Per1995} for further background and details. Let
$|\Phi\kt$ be any bipartite pure state of the tensor product space
${\cal H}^{A} \otimes {\cal H}^{B}$. It can be represented as
\beq
\label{eq:bipartstate}
|\Phi\kt = \sum_{i=1}^{N_A} \sum_{j=1}^{N_B} c_{ij} |ij\kt,
\eeq
where $\{|i\kt\}, \{|j\kt \}$ are
mutually orthonormal in their respective subspaces ${\cal H}^A$ and
${\cal H}^B$.

The \emph{reduced density matrices}
\beq
\rhoA=\tr_B\left(|\Phi\kt \br\Phi|\right), \;
\qquad
\rhoB=\tr_A\left(|\Phi\kt \br\Phi|\right),
\eeq
obtained after partially tracing the other subsystem is the state accessible to
either $A$ or $B$ respectively. They can be written in terms of the matrix $C$
whose elements are the coefficients $c_{ij}$ of the state
as $\rho_A=CC^{\dagger}$ and $\rho_B=(C^{\dagger}C)^T$
(where $A^T$ is the transpose of $A$).
These are evidently positive semi-definite matrices, and
let their eigenvalue equations be $\rhoA|\phi^A_j\kt =\lambda_j |\phi^A_j\kt$
and $\rhoB|\phi^B_j\kt =\lambda_j |\phi^B_j\kt$, with non-vanishing eigenvalues
indexed by $1 \le j \le N_A$.  The \emph{(Schmidt) eigenvalues} $\{\lambda_j\}$
of $\rhoA$ and $\rhoB$ are identical, except the larger subsystem ($B$) is
additionally padded with $N_B-N_A $ zero eigenvalues.  Additionally, assume
that the $\{\lambda_j\}$ are ordered such that
$\lambda_1 \ge \cdots \ge \lambda_{N_A}$.

The \emph{Schmidt decomposition}
\beq
\label{eq:DefnSchmidt}
|\Phi\kt= \sum_{j=1}^{N_A} \sqrt{\lambda_j} \, |\phi^A_j \kt |\phi^B_j\kt,
\eeq
is the most compact form of writing the bipartite state $|\Phi\kt$
in a product basis from orthonormal sets
$\{|\phi^A_j\kt\}$ and $ \{|\phi^B_j\kt\}$,
and uses the eigenvalues $\lambda_j$ and corresponding eigenvectors.
By normalization of the state $|\Phi\kt$ one has
$\sum_{j=1}^{N_A} \lambda_j = 1$.
The Schmidt decomposition follows from
the singular value decomposition of a matrix whose entries are
the coefficients of the state $|\Phi\kt $ in any product basis.
The state is unentangled if and only if $\lambda_1=1$ (and hence all other
eigenvalues are $0$), and the Schmidt decomposition gives the states of the
individual subsystems. Otherwise $\lambda_2>0$ and the Schmidt decomposition
consists of at least two terms.  For maximally entangled states,
$\lambda_j=1/N_A$ for all $j$.

Additionally the {\it closest product state} to $|\Phi\kt$ (in any metric
equivalent to the Euclidean) is $|\phi^A_1 \kt |\phi^B_1\kt$.  Hence the
largest eigenvalue of the reduced density matrices $\lambda_1$ is the maximum
possible overlap of the bipartite state with a product state. This provides a
geometric meaning to the Schmidt decomposition. A complementary question is the
one of identifying the {\it closest maximally entangled state} and the distance
to it.  Indeed the Schmidt decomposition is also the crucial ingredient in
answering this.  To the best of our knowledge this is not discussed in
introductions to entanglement, and hence will be addressed in more detail in
Sect.~\ref{sec:dist-max-entangled}.

The entanglement entropy in the state $|\Phi\kt$ is the
\emph{von Neumann entropy} of the reduced density matrices,
\begin{equation}
\begin{split}
S_1 &=-\tr\left (\rhoA \ln \rhoA \right)=-\tr\left (\rhoB \ln \rhoB \right)\\
    &=- \sum_{i=1}^{N_A} \lambda_i \ln \lambda_i.
\end{split}
\end{equation}
Thus if $S_1=0$, then the state is unentangled,
whereas a maximally entangled state has $S_1 = \ln N_A$.
More generally, to characterize entanglement one considers the moments
\beq
\label{eq:momentsdefn}
 \mu_{\alpha}= \sum_{i=1}^{N_A} \lambda_i^{\alpha}, \;\; \alpha >0.
\eeq
Normalization of the state $|\Phi\kt$, implies normalization of the
reduced density matrices:
$\mu_1=\tr(\rhoA)=\tr(\rhoB)=\sum_{i=1}^{N_A} \lambda_i =1$.
The second moment $\mu_2$ is the {\it purity} of the reduced density
matrices $\rhoA$ or $\rhoB$. We will also be especially interested
in the moment $\mu_{1/2}$ due to its connection with the
distance to the closest maximally entangled state,
as discussed ahead in Sect.~\ref{sec:dist-max-entangled}.

As the set of Schmidt eigenvalues $\{ \lambda_i, \, 1 \le i \le N_A \}$ defines
a classical probability measure, entropies can be defined based on the many
measures studied in this context.
The so-called
\emph{Havrda-Charv{\'a}t-Tsallis (HCT) entropies}~\cite{HavCha1967,Tsa1988,BenZyc2006} are:
\begin{equation}
  S_{\alpha} = \dfrac{1-\mu_{\alpha}}{\alpha-1},
\label{eq:Tsallis}
\end{equation}
while the \emph{R\'enyi entropies} \cite{Ren1961wcrossref} are
defined by
\begin{equation*}
  R_\alpha = \dfrac{\ln \mu_{\alpha}}{1-\alpha} .
\end{equation*}
The R\'enyi entropies are evidently additive, that is the entropy of
independent processes are sums of entropies of the individual processes,
whereas the HCT entropies are not. We will use the HCT entropies as the
ensemble averages are more easily done with $\mu_{\alpha}$ rather than with
$\ln \mu_{\alpha}$.  Both types of entropies limit to the von Neumann entropy
$S_1$ as $\alpha \rightarrow 1$.  Moreover, the \emph{purity} $\mu_2$ is
directly related the so-called \emph{linear entropy} $S_2=1-\mu_2$, and is
often used as a simpler measure of entanglement than the von Neumann
entropy. The state $|\Phi \rangle$ is unentangled if and only if the reduced
density matrices are pure, in which case all $\mu_\alpha=1$, or equivalently
$S_{\alpha}=0$, for $\alpha > 0$.

Let $|\Psi\kt$ be a random state, i.e.\ it is chosen at random uniformly with
respect to the Haar measure from the Hilbert space
${\cal H}^{A} \otimes {\cal H}^{B}$, this induces a probability density on the
eigenvalues $\{\lambda_i\}$. For our purposes it suffices to state that the
asymptotic (large $N_A$ and $N_B$ with fixed ratio $Q=N_B/N_A \ge 1$) limit
of the density of the scaled eigenvalues,
$\tilde{\lambda}_i=\lambda_i \, N_A$ is given by the \Marcenko-Pastur
distribution \cite{MarPas1967} as shown in \cite{SomZyc2004}
\beq \label{eq:marcenko-pastur-law}
\rho_{\text{MP}}^{Q}(x)
  = \frac{Q}{2\pi} \frac{\sqrt{(x_{+}-x)(x-x_{-})}}{x},
     \; x_{-} \le x \le x_{+},
\eeq
and $0$ otherwise. The distribution is in the finite support $[x_{-},x_{+}]$
where
\beq
x_{\pm}=1+\frac{1}{Q} \pm \frac{2}{\sqrt{Q}}.
\eeq

For quantum chaotic eigenfunctions the eigenvalues of the reduced density
matrix have been verified to follow $\rho_{\text{MP}}^Q(x)$
\cite{BanLak2002}. Detailed analysis, including exact results for finite $N_A$
are given in \cite{KubAdaTod2008,KubAdaTod2013}. Using
Eq.~\eqref{eq:marcenko-pastur-law}, the Haar averaged entanglement entropy is
\beq
\label{eq:Haar-S1-entanglement}
\begin{split}
\overline{S_1}
   &=- \sum_{i=1}^{N_A} \overline{\frac{\tilde{\lambda}_i}{N_A}
                                  \ln \frac{\tilde{\lambda}_i}{N_A} }
   =\ln N_A - \frac{1}{N_A} \sum_{i=1}^{N_A}\overline{ \tilde{\lambda}_i \ln \tilde{\lambda}_i }\\
   &= \ln N_A -\int_{x_{-}}^{x_{+} }x\ln x\ \rho_{\text{MP}}^{Q}(x) \, dx\\
   &= \ln N_A -\frac{1}{2Q}.
\end{split}
\eeq

While this is the large $N_A$ result, exact finite $N_A$ results are remarkably
enough known \cite{Pag1993, Sen1996}.  Equation~\eqref{eq:Haar-S1-entanglement}
seems to indicate that typical random states are almost as entangled as the
maximum possible $S_1 = \ln N_A$.
Similarly,  using Eq.~\eqref{eq:marcenko-pastur-law} one gets
\beq \label{eq:S-2-Haar}
\begin{split}
\overline{S_2}
   &=1- \sum_{i=1}^{N_A}\overline{\frac{\tilde{\lambda}_i^2}{N_A^2} }
   = 1-\frac{1}{N_A} \int_0^{4} x^2 \,\rho_{\text{MP}}^{Q}(x) dx\\
   &= 1-\frac{Q+1}{ N_A\, Q}.
\end{split}
\eeq
and the exact finite--$N_A$ result is \cite{Lub1978}
\begin{equation}
 \overline{S_2 } = 1-\frac{N_A + N_B}{1+ N_A N_B}.
\end{equation}
For the most part our numerical results will be for the
symmetric case $N_A=N_B$, corresponding to $Q=1$.

\subsection{Distance to the closest maximally entangled state}
\label{sec:dist-max-entangled}

Any state of ${\cal H}^{A} \otimes {\cal H}^{B} $ having the form
\beq
\frac{1}{\sqrt{N_A}} \sum_{i=1}^{N_A}  \sum_{j=1}^{N_B} u_{ij} |ij\kt,
\eeq
where  $U=\{ u_{ij} \}$ is a (generally) rectangular array such that
\beq
\label{eq:uudager}
U U^{\dagger}=\mathds{1}_{N_A}
\eeq
is maximally entangled. This follows as the reduced density matrix $\rho_A$
is then the most mixed state $\mathds{1}_{N_A}/N_A$,
corresponding to $\lambda_i=1/\sqrt{N_A}$.
Therefore, finding the closest maximally entangled state to
a state $|\Phi\kt$ as given in Eq.~(\ref{eq:bipartstate}) requires finding the
closest such array to the matrix $C'= \sqrt{N_A} C$, where $C$ is the array
of coefficients $c_{ij}$ as in Eq.~(\ref{eq:bipartstate}).
In the symmetric case this reduces to finding the closest unitary matrix
to a given one, a problem dealt with in Ref.~\cite{Kel1975}.
We provide an alternate proof and generalize to the case of
a rectangular array.

Let $C'$ be an arbitrary $N_A \times N_B$ matrix.
The task is to find another matrix $U$ of
the same shape satisfying Eq.~(\ref{eq:uudager})
and which minimises $\|C' - U \|^2$.
This minimization is equivalent to maximization  of $\Re
\tr(C' \, U^\dagger)$ over $U$ as the first and third
terms in the expansion $\|C' - U \|^2 =
\tr(C' C'^\dagger) - 2\, \Re\tr(C' \,U^\dagger) + N$ are
constant. Let the singular value decomposition of $C'$ be $V_1 \sqrt{S} V_2^T$,
where $V_1$ and $V_2$ are $N_A$ and $N_B$ dimensional unitary matrices
respectively, and $\sqrt{S}$ is a $N_A \times N_B$ dimensional
``diagonal" matrix with entries $\sqrt{s_i} \ge 0$ only when the
row and column indices are the same ($=i$) and are zero elsewhere.

The following holds:
\begin{equation}
\begin{split}
  \Re \tr(C' U^\dagger)
   & = \Re \tr(V_1 \sqrt{S} V_2^T U^\dagger)\\
   & = \Re \tr(\sqrt{S} V_2^T U^\dagger V_1).
\end{split}
\end{equation}
Note that $\tilde{U}= V_2^T U^\dagger V_1$ is an $N_B \times N_A$ array
such that $\tilde{U}^{\dagger} \tilde{U}=\mathds{1}_A$, and  $\Re \tr
(\sqrt{S} \tilde{U}) = \sum_{i=1}^{N_A} \sqrt{s_i} \, \Re \tilde{U}_{ii}$.
Now as $\sqrt{s_i}\geq 0$, $\Re \tr (\sqrt{S} \tilde{U})$
will be the maximum for
any $\tilde{U}$ having $\Re \tilde{U}_{ii}=1$ for all $i\leq N_A$,
where it is defined.
As $\tilde{U}$ elements are such that
$\sum_{i=1}^{N_B} |\tilde{U}_{ij}|^2 = 1$ for all $j \le N_A$, hence
the only array with $\Re \tilde{U}_{ii}=1$ for all $i$ is the
rectangular ``identity" matrix, that is
$\tilde{U}_{ij}=\delta_{ij}$ for $1\leq i \leq N_A$.
Therefore the closest required array $U$,
such that $UU^{\dagger}=\mathds{1}_{N_A}$, to the matrix $C'$
is $U_* = V_1 \tilde{U}^{\dagger} V_2^T$, which is essentially the
product of the two unitary matrices in the singular value decomposition of $A$.
Thus the closest maximally entangled state to $|\Phi\kt$ is
\beq
\label{eq:closeststate}
\begin{split}
|\Phi_*\kt
& = \frac{1}{\sqrt{N_A}} \sum_{i=1}^{N_A} \sum_{i=1}^{N_B} \sum_k(V_1)_{ik} (V_2^{T})_{kj}|ij\kt \\
& =\frac{1}{\sqrt{N_A}} \sum_{k=1}^{N_B} \sum_{i=1}^{N_A} (V_1)_{ik}|i\kt \sum_{j=1}^{N_B} (V_2)_{jk}|j\kt\\
&= \frac{1}{\sqrt{N_A}}\sum_{k=1}^{N_A}|\phi_k^A \kt |\phi_k^B\kt.
\end{split}
\eeq
Here $|\phi_j^{A}\kt$
($|\phi_j^{B}\kt$) are the eigenvectors of $\rho^{A}$ ($\rho^{B}$), i.e.\
Schmidt eigenvectors, with the pair having common index $j$ being
chosen to have the same eigenvalue $\lambda_j$.

The norm chosen to measure the distance $d_*$ is not important,
as long as it is a unitarily invariant one.  A good choice is given by,
\beq
\label{eq:closeststate2}
d_*^2 = \| |\Phi\kt - |\Phi_*\kt \|^2
      = 2 \left( 1 - \frac{1}{\sqrt{N_A}}\sum_{j=1}^{N_A}
        \sqrt{\lambda_j}\right),
\eeq
where $\| |\phi\kt \|=\sqrt{\br \phi|\phi \kt}$ is the Euclidean norm.
For an unentangled (product) state $\lambda_j =\delta_{1j}$, and the distance
$d_*$ is the largest possible,
\begin{equation} \label{eq:dstar-unentangled-product}
   d_*^{\text{product} }= \sqrt{2(1 - 1/\sqrt{N_A})},
\end{equation}
which for $N_A\to \infty $ converges to $\sqrt{2}$.
For a random state in the $N_A \times N_B$ space the mean squared distance
$\overline{d_*^2}$ can be calculated in the limit of large dimensionalities as
\begin{align}
\label{eq:dstar}
&\overline{d_*^2}_{\text{RMT}}(Q)
      =2\left( 1- \int_{x_{-}}^{x_{+}} \sqrt{x} \, \rho_{\text{MP}}^{Q}(x)\,dx \right) =\\
&2 -\frac{4}{3 \pi} \left(1+\frac{1}{\sqrt{Q}}\right) \left[ \left(Q+1\right)E(\kappa)-\left(\sqrt{Q}-1\right)^2 K(\kappa)\right] \nonumber,
\end{align}
where $\kappa=2/(Q^{1/4}+Q^{-1/4})$ and $E(\kappa)$ and $K(\kappa)$
are complete Elliptic integrals of the second and
first kind respectively \cite{GraRyz2014}.

For the symmetric case, $Q=1$, the above simplifies to
\begin{equation}
\overline{d_*^2}_{\text{RMT}}= 2\left( 1 -\frac{8}{3\pi} \right) \approx 0.302,
\label{eq:dstar-random}
\end{equation}
where we drop the explicit specification of $Q$.
Thus for a typical random state in a  symmetric setting
$\sqrt{\overline{d_*^2}_{\text{RMT}}}/d_*^{\text{product}} \approx \sqrt{1-8/(3 \pi)} \approx 0.388$.
This indicates that whereas for typical random states the entanglement entropy,
Eq.~\eqref{eq:Haar-S1-entanglement} is nearly maximal, the states themselves
are quite far from being maximally entangled.  Detailed results about
typicality of these distances, and their relationship to the negativity measure
of entanglement can be found in~\cite{BuSinZhaWu2016:p}. Perturbation theory
and numerical results for this quantity are presented
in Sect.~\ref{sec:moment-half}; see Fig.~\ref{fig:dstar}.

In the asymmetric cases, $\overline{d_*^2}_{\text{RMT}}(Q)$
decreases monotonically as $Q$ increases from $1$ to $\infty$
and vanishes for large $Q$ as
\beq
\overline{d_*^2}_{\text{RMT}}(Q) \sim \frac{1}{4Q}.
\eeq
Thus as the ``environment" (the larger subsystem) grows relatively in size,
typical states are not only highly entangled, but are also metrically close to
maximally entangled states. Thus if dynamics drives states close to these
typical states one may say that they would thermalize and reach the ``infinite
temperature" ensemble of Floquet nonintegrable systems
\cite{LazDasMoess2014pre,LucaRigol2014}.

\section{Universal entanglement transition}

The derivation of the transition in the eigenstate entanglement from
unentangled to that typical of random states begins with the introduction of a
random matrix model and application of standard Rayleigh-Schr\"odinger
perturbation theory.  From these expressions, the transition parameter can be
identified.  Then as soon as one attempts to apply ensemble averaging to the
resultant expressions, it becomes necessary to regularize appropriately the
perturbation theory to account properly for small energy denominators.
Finally, it turns out to be possible in this case to go beyond the perturbative
regime for the entanglement entropies by a recursive embedding of the
regularized perturbation theory that leads to a simple differential equation
which is straightforward to solve.

\subsection{Random matrix transition ensemble}

Random matrix models for breaking fundamental or dynamical symmetries have been
introduced for a variety of cases since the work of~\cite{PanMeh1983} for
breaking time reversal invariance.  Examples include ensembles to describe
parity violation~\cite{TomJohHayBow2000}, parametric statistical
correlations~\cite{GolSmiBerSchWunZel1991}, modeling transport
barriers~\cite{BohTomUll1993,TomUll1994,MicBaeKetStoTom2012}, and the
fidelity~\cite{CerTom2003}.  The structure imposed by the particular symmetry
is incorporated into the unperturbed ensemble (zero$^{th}$-order piece) and a
tunable-strength ensemble is added that violates that structure.  Each symmetry
is different, and for the case of non-interacting, strongly chaotic subsystems,
the direct product structure must be imposed on the zero$^{th}$-order part of
the ensemble, which is violated by an interaction part not respecting that
structure.  To model the statistical behavior of bipartite systems, such as
those given by Eq.~(\ref{eq:Hab}) or Eq.~\eqref{eq:Uab}, a \emph{random matrix
  transition ensemble} has been introduced recently in
Ref.~\cite{SriTomLakKetBae2016},
\begin{equation}
\label{eq:mod}
  {\cal U}_{\text{RMT}}(\epsilon) =
      \left( U_{A}^{\text{CUE}} \otimes U_{B}^{\text{CUE}}\right)
      \,  U_{AB}(\epsilon),
\end{equation}
where the tensor product is taken of two independently
chosen members of the circular unitary
ensemble (CUE) of dimension $N_A,N_B$, respectively,
and $U_{AB}(\epsilon)$ is a diagonal unitary matrix in the resulting
$N_AN_B$--dimensional space representing the coupling.
Its diagonal elements are taken as $\exp(2 \pi i \epsilon \, \xi_j)$
where $\xi_j$ are independent random variables that are uniform on the
interval $(-1/2,1/2]$.
Preparing for the perturbation theory introduced ahead,
it is helpful to define a Hermitian matrix $V_{AB}$ such that
\beq
\label{vjj}
  U_{AB}(\epsilon)=\exp(i\epsilon V_{AB})
  \quad {\rm where}\ \ (V_{AB})_{jk}=2\pi\xi_j \delta_{jk} .
\eeq
The strength of the coupling $\epsilon$ is a real number, and
$1\le j,k \le N_AN_B$ label the basis states of the subsystem.  A limiting case
of this ensemble has been studied previously~\cite{LakPucZyc2014}, wherein the
entangling power of ${\cal U}_{\text{RMT}}(\epsilon=1)$ was found
analytically. If $\epsilon=0$ there is no coupling between the spaces and there
is no eigenstate entanglement, and the consecutive neighbor spacing statistics
is Poissonian \cite{TkoSmaKusZeiZyc2012,SriTomLakKetBae2016}.  As $\epsilon$ is
increased, it leads to a rapid transition in the spacing statistics to the
Gaussian or circular unitary ensemble, and the entanglement also reaches values
that are valid for random states in the entire $N_AN_B$--dimensional Hilbert
space \cite{SriTomLakKetBae2016,LakSriKetBaeTom2016}.

\subsection{Applying Rayleigh-Schr\"odinger perturbation theory}

Perturbation theory for random matrix ensembles has previously been applied to
describe symmetry
breaking~\cite{FreKotPanTom1988,Tom1986,TomUll1994,TomJohHayBow2000},
and
continues to be of interest due to various applications ranging from quantum
mechanics to quantitative finance~\cite{LeySel1990,AllBou2012,BenEnrMic2017}.
Mostly this has been done in a Hamiltonian framework whereas the ensemble of
Eq.~\eqref{eq:mod} is unitary for which the spectrum lies on the unit circle in
the complex plane.  The first and second order corrections are given in
App.~\ref{app:perturbation} for the eigenvalues and eigenvectors of
$\mathcal{U}(\epsilon)$ of Eq.~\eqref{eq:Uab}.  In the limit of
$N_A\rightarrow \infty$, the local part of the spectrum of relevance to
perturbation theory occupies a differentially small fraction of the unit circle
and it is straightforward to expand the perturbation theory for the unitary
ensemble in order to make it look just like the standard perturbation
expressions for Hamiltonian systems with the use of Eq.~\eqref{vjj}; locally
the correlations built into the unitary matrix elements can be ignored.  Thus,
corrections of $\mathcal{O}(N_A^{-1})$ are to be ignored from the outset in the
derivation presented in this subsection.

For each single member of the ensemble ${\cal U}_{\text{RMT}}(\epsilon)$, the
basis chosen with which to apply perturbation theory is given by the
eigenstates of a single realization of
$U_{A}^{\text{CUE}} \otimes U_{B}^{\text{CUE}}$.  Denote the set by
$|j^A\kt |k^B \kt$, where $|j ^{A}\kt$ and $|j^{B}\kt$ are eigenstates of
individual members $U_{A}$ and $U_B$, respectively.  In the following the
labels $A,B$ are dropped from this basis as the ordering implies the particular
subsystem, i.e.\ $|j^A\kt |k^B \kt \equiv |j k \kt$.  The labels AB are also
dropped from $U_{AB}(\epsilon)$ and $V_{AB}$ as well for convenience.  The
eigenstates of the ensemble of ${\cal U}_{\text{RMT}}(\epsilon=0)$ are
unentangled, and uniformly random with respect to the direct product Haar
measures of the subspaces.

\begin{figure}[b]
\includegraphics{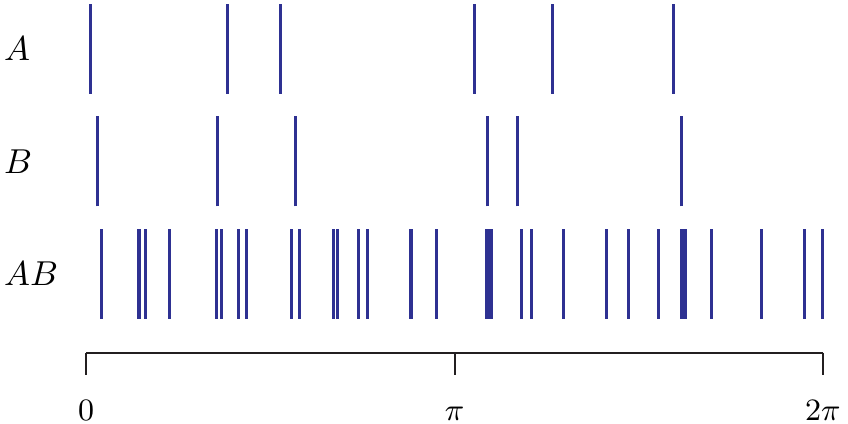}
\caption{\label{fig:phases}%
Example spectrum of the direct product of two uncoupled subsystems,
Eq.~\eqref{specab}, for $N_A=N_B=6$.  The spectra are ordered vertically
for subspaces $A$, $B$, and the combined spectrum, respectively.
This schematically illustrates two key features,
the emergence of Poissonian spacing fluctuations and the great
increase in eigenangle density for the combined spectrum.}
\end{figure}

The associated eigenangles are given by
\beq
\label{specab}
 \theta_{jk}=\theta_j+\theta_k \quad {\rm mod}(2\pi)  ,
\eeq
see Fig.~\ref{fig:phases}.  Note that the mean spacing of eigenangles for
$\theta_j$ ($\theta_k$) is $2\pi/N_A$ ($2\pi/N_B$), and for $\theta_{jk}$ is
$2\pi/N_AN_B$, showing that the combined spectrum is denser by a factor of
either $N_A$ or $N_B$ than the individual subspace spectra.  Choose a
particular eigenstate of an ensemble member ($\epsilon\ne 0$), denoted
$|\Phi_{j k}\kt $, as the one which is continuously connected to $|jk\kt$ as
$\epsilon$ vanishes.  Its Schmidt decomposition can be written as
\beq
\label{eq:Schmidt}
|\Phi_{j k}\kt= \sum_{j=1}^{N_A} \sqrt{\lambda_j} |\phi^A_j \kt |\phi^B_j\kt,
\eeq
where $\lambda_1\ge \lambda_2 \ge \cdots \ge \lambda_{N_A}$ are the (Schmidt)
eigenvalues of the reduced density matrix $\rho^{A}$, and the states
$\{|\phi^A_j \kt, \;|\phi^B_j\kt\}$ are the orthonormal eigenvectors of $\rhoA$
and $\rhoB$ respectively.  The perturbation expression to first order of the
eigenstate is (${\cal U}(\epsilon)$ is the member of the ensemble
${\cal U}_{\text{RMT}}(\epsilon)$ under consideration with corresponding $V$)
\beq
\label{pertpure}
|\Phi_{j k} \kt \approx |jk \kt + \epsilon  \sum_{j^\prime k^\prime \neq j  k}
     \dfrac{\br j^\prime k^\prime |V |jk\kt}
           {\theta_{jk}-\theta_{j^\prime k^\prime}}
      \, |j^\prime k^\prime \kt \ .
\eeq
where here and throughout $j^\prime k^\prime\ne jk$ means to exclude the single
term in which both $j^\prime=j$ and $k^\prime=k$.  Note that the eigenbasis
$|jk\kt$ of one member of $U_{A}^{\text{CUE}} \otimes U_{B}^{\text{CUE}}$ is
not the one in which the operator $V$ is diagonal, and a transformation to the
appropriate eigenbasis must be performed in order to evaluate the matrix
elements above.  Due to the statistical properties of the ensemble, the
transformation between bases is statistically equivalent to choosing a direct
product random unitary transformation independently uniform with respect to the
Haar measure in each subspace.  Thus, there is a central limit theorem for the
behavior of the matrix elements.  Furthermore, there is a complete absence of
correlations between the unperturbed spectra and the unperturbed eigenstates.
Like the classical random matrix ensembles, there is an
ergodicity~\cite{Pan1979} for this ensemble as well in that spectral averages
taken over an individual member of the ensemble in the large dimensional limit
has fluctuation properties that are nearly equal those expected of the
ensemble.

An interesting circumstance appears when calculating the density matrix from
the perturbation expression, only one row and column have matrix elements of
$\mathcal{O}(\epsilon)$,
\begin{eqnarray}
\rhoA &=& |j \kt \br j| + \epsilon \sum_{j^\prime \ne j} \frac{\langle j^\prime k|V |jk\rangle}{\theta_{j}-\theta_{j^\prime}}|j^\prime \rangle \langle j |+h.c. \nonumber \\
&& +\epsilon^2  \sum_{j^\prime k^\prime ,j^{\prime\prime} k^\prime \neq j  k} \dfrac{\br jk |V |j^{\prime\prime} k^\prime \kt\br j^\prime k^\prime |V |jk \kt}{\left(\theta_{jk}-\theta_{j^\prime k^\prime }\right)\left(\theta_{jk}-\theta_{j^{\prime\prime} k^\prime}\right)} \, |j^\prime \kt\br j^{\prime\prime} |.  \nonumber \\
\end{eqnarray}
However, given the comments above about the statistical nature of the ensemble
members, matrix elements vary as zero centered Gaussian random variables
without correlation to the energy denominators.  In fact, the dominant terms
are nearly always those with the smallest energy denominators.  The important
point is that the eigenangle denominators of the $\mathcal{O}(\epsilon)$ terms
involve only the spectrum of subsystem $A$, whose differences are generally a
factor $N_B$ greater ($N_B \ge N_A \rightarrow \infty$) than energy
denominators involving the full spectrum; recall Fig.~\ref{fig:phases}.  So
these terms represent an $\mathcal{O}(N_B^{-1})$ correction and can be dropped.

Thus, it is necessary to self-consistently track down all the
$\mathcal{O}(\epsilon^2)$ perturbations contributing to the reduced density
matrix.  It turns out that all the second order terms not shown in
Eq.~\eqref{pertpure} contributing to the summation over the $|jk\kt$ states
contribute at order $\mathcal{O}(\epsilon^3)$ or greater and can be dropped.
The only second order correction that must be accounted for in
Eq.~\eqref{pertpure} is the second order normalization of the $|jk\kt$ term.
Thus, the normalized state with all the terms needed to this order
is~\cite{SakTua1994}
\beq
\label{eq:2pert}
\begin{split}
|\Phi_{j k}\kt
 = & \left( 1- \frac{\eps^2}{2} \sum_{j^\prime k^\prime \neq j k}
         \dfrac{|\br jk|V |j^\prime k^\prime \kt|^2}{(\theta_{jk}-\theta_{j^\prime k^\prime})^2} \right)
   |j k\kt\\
  & + \eps \sum_{j^\prime k^\prime \neq j  k}
    \dfrac{\br j^\prime k^\prime |V |jk\kt}{\theta_{jk}-\theta_{j^\prime k^\prime}}
           |j^\prime k^\prime \kt.
\end{split}
\eeq
The expression for the density matrix is finally,
\begin{eqnarray}
\label{dmff}
\rhoA &=& \left( 1- \eps^2 \sum_{j^\prime k^\prime \neq j k}
         \dfrac{|\br jk|V |j^\prime k^\prime \kt|^2}{(\theta_{jk}-\theta_{j^\prime k^\prime })^2} \right) |j \kt \br j| \nonumber \\
&&  +\epsilon^2  \sum_{j^\prime k^\prime ,j^{\prime\prime} k^\prime \neq j  k} \dfrac{\br jk |V |j^{\prime\prime} k^\prime \kt\br j^\prime k^\prime |V |jk \kt}{\left(\theta_{jk}-\theta_{j^\prime k^\prime }\right)\left(\theta_{jk}-\theta_{j^{\prime\prime} k^\prime}\right)} \, |j^\prime \kt\br j^{\prime\prime} |.  \nonumber \\
\end{eqnarray}

\subsection{Schmidt decomposition}

Given the eigenstate expressions for the ensemble, the next step towards
evaluating the entanglement entropies is to deduce the Schmidt eigenvalues.  A
priori, the perturbed state in Eq.~(\ref{eq:2pert}) is not in the Schmidt
decomposed form of Eq.~(\ref{eq:Schmidt}) or alternatively, the density matrix
expression in Eq.~\eqref{dmff} has not yet been diagonalized. Therefore it is
not yet clear what the reduced density matrix eigenvalues are from the
expressions.  However, the ensemble of Eq.~\eqref{eq:mod} has some very nice
statistical properties, as mentioned in the previous subsection, that greatly
simplify the process.

The key argument is that, given the random, featureless interaction (all matrix
elements fluctuate about the same scale, independent of the pair of states
involved), the dominant terms in perturbation
theory come from the nearest neighbors in the spectrum.
This has the consequence of
inhibiting any changes to the Schmidt eigenvectors.  Consider the difference of
$\theta_{j+1}-\theta_{j}$, which would result if the value of $k$ is
unchanged. It is likely to be much farther away than the nearest neighbors.
Much more likely is that an appropriate change in $k$ can help cancel a large
part of this difference.  For example, $|j+1,k\kt$ and $|j,k+1\kt$ has a far
greater chance of being nearby as
$\theta_{j+1}+\theta_{k}-\theta_{j}-\theta_{k+1}$ has a far greater chance of
generating a small difference.  In fact, for $N_A \rightarrow \infty$ in a
local neighborhood of the spectrum, for any given value of $j$, there is only
one value of $k$ that has a chance of being close.  Similarly, for a given
value of $k$, only one value of $j$ has a chance of being close.  Thus, the
only terms that are appreciable in Eq.~\eqref{eq:2pert} lead to mutually
orthonormal sets of $\{j,k\}$ locally in the spectrum.  The same argument
implies that Eq.~\eqref{dmff} is diagonal as well.  Since $j,j^\prime$ are
matched to the same values of $k$, and a unique value of $j$ gives a near
neighbor, one of the diagonal terms will nearly always dominate the
$j\ne j^\prime$ terms.  Thus as a consequence, under these conditions, we have
with high probability that the perturbed state in Eq.~(\ref{eq:2pert}) remains
in the Schmidt form to an excellent approximation that gets better with
increasing dimensionality.  It turns out that corrections to this picture
produce higher order corrections after ensemble averaging.  Thus, the
eigenvalues of the reduced density matrix can be taken directly from the
diagonal elements of Eq.~\eqref{dmff}.

\subsection{Eigenvalues of the reduced density matrix and ensemble averaging}
\label{erdmea}

Following from Eq.~\eqref{dmff}
the expression for the largest eigenvalue $\lambda_{1}$
of the reduced density matrix  $\rhoA$ is approximately
\begin{equation}
\label{eq:lambda1}
\lambda_1 \approx  1- \eps^2 \sum_{j^\prime k^\prime \neq j k}
         \dfrac{|\br jk|V |j^\prime k^\prime \kt|^2}{(\theta_{jk}-\theta_{j^\prime k^\prime})^2}
\end{equation}
and the second largest is nearly always
\beq
\label{eq:lambda2}
\lambda_2 \approx  \eps^2 \dfrac{|\br jk |V |j_1 k_1\kt|^2}{\left(\theta_{jk}-\theta_{j_1k_1}\right)^2},
\eeq
where  $\theta_{j_1k_1}$ is the closest eigenangle to $\theta_{jk}$.
The dependence of these eigenvalues on the ``central state" indices $jk$ is
suppressed.  To this level of approximation, the largest eigenvalue is simply
the squared overlap of the state with its perturbation.  Such a situation has
been analyzed in the context of parametric eigenstate correlators and the
fidelity with a broad variety of physical
applications~\cite{BruLewMuc1996,AlhAtt1995,AttAlh1995,KusMit1996,CerTom2003}.

The unperturbed Schmidt vector corresponding to the largest eigenvalue $| j\kt$
is unchanged in this approximation.  This is a statistical approximation that
simply reads off the eigenvalues of the reduced density matrix assuming that
the significant terms in the perturbation expansion in Eq.~(\ref{eq:2pert}) are
determined by eigenangle denominators and relies on the $|j^\prime \kt$ vectors
all being locally orthonormal, and likewise the $|k^\prime \kt$ vectors.  A
more detailed analysis is possible~\cite{Lakshminarayan19} that entails
constructing the reduced density matrix and making a further perturbation
approximation to diagonalize it. This leads to some changes in the formal
structure of these expressions.  For example, the sum in the largest eigenvalue
$\lambda_1$ (Eq.~(\ref{eq:lambda1})) is changed to a more restricted sum such
that $j^\prime \ne j$ {\it and} $k^\prime \ne k$.  However, these changes only
lead to higher order corrections after ensemble averaging and do not affect any
of the results presented here.

In order to prepare these expressions for ensemble averaging ahead, it is quite
useful to extract certain scales and introduce spectral two-point functions,
which can be used to express the summations in integral
form~\cite{FreKotPanTom1988}.  First, the statistical properties of the
transition matrix elements and the energy differences each involve a single
scale.  The admixing matrix elements
$\epsilon^2 |\br jk|V |j^\prime k^\prime \kt|^2$
in the unperturbed basis behave as the absolute square
of a complex Gaussian random variable.  Denoting the mean as $v^2$, the term
$\epsilon^2 |\br jk|V |j^\prime k^\prime \kt|^2$ is replaced by
$v^2 w_{j^\prime k^\prime}$, where the set of $\{w_{j^\prime k^\prime}\}$
behave according to the statistical probability density
\beq
\label{eq:off-diagonal-matrix-elements}
\rho(w)=e^{-w}, \;\; w \ge 0,
\eeq
with unit mean across the ensemble.  This is a property of the ensemble,
Eq.~\eqref{eq:mod}.
Note that for a bipartite chaotic dynamical system with the structure of
Eq.~\eqref{eq:Uab}, for which the perturbation is not random,
the distribution of $w$ shows deviations
from the exponential as illustrated in App.~\ref{app:distribution-of-w}.
Still the predictions of the random matrix ensemble
apply very well to the coupled kicked rotors, as will be seen below.

Secondly, the energy denominator has the single scale of a mean spacing,
$D=2\pi/(N_AN_B)$, built in.
Let $\theta_{jk}-\theta_{j^\prime k^\prime}=Ds_{j^\prime k^\prime }$, upon substitution,
\begin{equation}
\label{eq:lambda1new}
\lambda_1 \approx  1 - \frac{\epsilon^2v^2}{D^2} \sum_{j^\prime k^\prime \neq j k}  \dfrac{w_{j^\prime k^\prime }}{s_{j^\prime k^\prime }^2}
= 1 - \Lambda \sum_{j^\prime k^\prime \neq j k}  \dfrac{w_{j^\prime k^\prime}}{s_{j^\prime k^\prime }^2},
\end{equation}
where the second form uses the definition
of the \emph{transition parameter}~\cite{PanMeh1983,FreKotPanTom1988},
\begin{equation} \label{eq:transition-parameter}
  \Lambda = \frac{\epsilon^2v^2}{D^2} .
\end{equation}
This illustrates its natural appearance in the perturbation expressions.

Let's define the function
\beq
\label{twopointws}
R(s, w) =  \sum_{j^\prime k^\prime \ne jk} \delta \left( w - w_{j^\prime k^\prime} \right) \delta \left( s - s_{j^\prime k^\prime} \right) .
\eeq
where $R(s, w)$ is the probability density of finding a level at a rescaled
distance $s$ from $\theta_{jk}$, and the corresponding matrix element variable
$w$ takes the value $w_{j^\prime k^\prime}$.  In general there could be
correlations between matrix elements and spacings, however for the current
scenario they are completely independent across the ensemble.
Equation~\eqref{eq:lambda1} can be expressed in an exact integral form using
$R(s,w)$ as,
\beq
\label{lambda1eq}
  \lambda_1 \approx
       1 - \Lambda \int^\infty_{-\infty}{\rm d}s
           \int_0^\infty {\rm d}w \frac{w}{s^2} R(s, w) .
\eeq
This expression clarifies a number of issues.  First, $\Lambda$ controls the
entanglement behavior as asserted in the introduction, and the perturbative
regime is for $\Lambda \ll 1$; see Sect.~\ref{sec:universal-scaling-parameter}.
Another great simplicity is that ensemble averaging of $\lambda_1$ is reduced
entirely to replacing $R(s, w) $ with its ensemble average,
$\overline{R(s, w) }$.  For the ensemble of Eq.~\eqref{eq:mod}, all
correlations between matrix elements and spacing vanish.  In fact,
\beq
\label{ensave}
\overline{R(s, w) } = \rho(w) R_2(s) = \exp(-w),
\eeq
where the last form follows for the ensemble because
$R_2(s)=1\ (-\infty \le s \le \infty)$ for a Poissonian sequence
of unit mean and in general defined as,
\beq
  R_2(s) = \overline{\sum_{j^\prime k^\prime \ne jk}
            \delta \left( s - s_{j^\prime k^\prime} \right)}.
\eeq
The eigenangles are pairwise sums of those with CUE fluctuations. These lose
correlations and are Poisson distributed. For a recent rigorous mathematical
proof in the context of product unitary operators
see~\cite{TkoSmaKusZeiZyc2012}.  Another issue is that of calculating moments
of the Schmidt eigenvalues, which is treated in Sect.~\ref{momentsrdm}.  We
only mention here that the moments involve more complicated expressions than
are given in Eq.~\eqref{ensave}.  Finally, the spacing integral in
Eq.~\eqref{lambda1eq} using Eq.~\eqref{ensave} for ensemble averaging leads to
a divergence which requires a regularization as there are too many small
spacings.  The regularization is given in Sect.~\ref{regularizations}.

A curious feature arises for the ensemble averaging of the expression for
$\lambda_2$.  The expression for the nearest neighbor spacing density of a
Poissonian sequence arises from the consideration of consecutive spacings
between eigenvalues.  However, $|j_1k_1\rangle$ is selected to be the closest
nearest neighbor.  In effect, $\theta_{jk}$ has a nearest neighbor to the left
and one to the right in the spectrum, and $|j_1k_1\rangle$ is determined by the
smaller spacing of the two.  The standard result $\rho_{\text{NN}}(s)=\exp(-s)$
is not the appropriate density.  Rather, the \emph{\CNtxt} density,
$\rho_\CN(s)$, is given by \beq \rho_\CN(s) = 2 \exp(-2s) \ \ (0\le s\le
\infty) \eeq for a Poissonian spectrum of unit mean
spacing~\cite{LakSriKetBaeTom2016,OurRNNS}.

Collecting results gives
\begin{subequations} \label{eq:lambda2-again}
\begin{eqnarray}
\label{eq:lambda2-again-a}
\overline{\lambda_1} &\approx& 1 - \Lambda \int^\infty_{-\infty}{\rm d}s \int_0^\infty {\rm d}w \frac{w}{s^2} e^{-w} \\
\overline{\lambda_2} &\approx &  2 \Lambda \int_0^{\infty}\frac{w}{s^2}e^{-w-2s}\, dw\, ds.
\label{eq:lambda2-again-b}
\end{eqnarray}
\end{subequations}
As mentioned earlier for $\lambda_1$, both equations~\eqref{eq:lambda2-again}
diverge as $s\rightarrow 0$.  The divergence indicates that the correct order
of $\overline{\lambda_1}$ or $\overline{\lambda_2}$ is not, in fact,
$1-\mathcal{O}(\Lambda)$ or $\mathcal{O}(\Lambda)$, respectively. Instead, they
turn out to be of the form
$\overline{\lambda_1} =1-\mathcal{O}(\sqrt{\Lambda})$ and
$\overline{\lambda_2} =\mathcal{O}(\sqrt{\Lambda})$, and indeed the average of
the general moments in Eq.~(\ref{eq:momentsdefn}) for $\alpha > 1/2$ differ
from $1$ by terms of the $\mathcal{O}(\sqrt{\Lambda})$, as shown in
Sect.~\ref{momentsrdm}.  Thus the eigenvalues and moments are much more
``volatile" than suggested by naive perturbation theory. This will be further
related in Sect.~\ref{sec:distribution-of-schmidt-eigenvalues} to the emergence
of power laws in the probability density of reduced density matrix eigenvalues
in this regime.

\subsection{The transition parameter $\Lambda$}
\label{sec:universal-scaling-parameter}

Once the universal transition curve is derived for the random matrix ensemble
\eqref{eq:mod}, it is of interest to explore how well the universal results
apply to chaotic dynamical systems.  For that reason, instead of deriving the
expression for $\Lambda$ exclusively for the ensemble of Eq.~\eqref{eq:mod}, a
slightly more general expression is derived.  The subsystem unitary operators
are taken from the CUE (imitated well by the chaotic dynamical subsystems), but
the interaction is left unspecified allowing for greater flexibility and
generality of applicability.  It turns out that the transition parameter
expression under such circumstances is given by (see
App.~\ref{app:universal-scaling-parameter} for a detailed derivation)
\begin{equation}
\begin{split}
& \Lambda=\dfrac{N_A^3 N_B^3}{4 \pi^2(N_A^2-1)(N_B^2-1)} \times \\
& \left(1+ \left|\frac{\tr U_{AB}}{N_A N_B}
  \right|^2  - \frac{1}{N_A}\, \left\|\frac{U^{(A)}}{N_B}\right\|^2
     -\frac{1}{N_B}\, \left\|\frac{U^{(B)}}{N_A}\right\|^2 \right).
\end{split}
\label{eq:lambda1-new}
\end{equation}
Here $ U^{(A)}_{ii}=\sum_{k}(U(\epsilon))_{ik}$,
$U^{(B)}_{kk}=\sum_{i}(U(\epsilon))_{ik}$ are partially traced (still diagonal)
interaction operators, which are in general not unitary, and
$\|X\|^2=\Trace(XX^{\dagger})$ is the Hilbert-Schmidt norm.  It can be checked
that this $\Lambda$ correctly vanishes if $V$ is a function of either
coordinate alone, and thereby not entangling. However, it does not vanish for
general separable potentials, and hence the assumption that {\it the
  interaction is entangling} is necessary.  Equation \eqref{eq:lambda1-new} is
the generalization of the corresponding result stated in
Ref.~\cite{SriTomLakKetBae2016} for the case $N_A=N_B=N$.

Performing an additional ensemble average over the diagonal random phases
for the $U(\epsilon)$ of Eq.~\eqref{eq:mod} gives
\begin{equation}
\begin{split}
\Lambda_{\text{RMT}} &= \dfrac{N_A^2 N_B^2}{4 \pi^2(N_A + 1)(N_B + 1)}
 \left[1-\dfrac{\sin^2 (\pi \epsilon)}{\pi^2\epsilon^2} \right]\\
 & \approx \dfrac{N_A N_B \epsilon^2}{12 },
\end{split}
\label{eq:lambdarmt}
\end{equation}
where for the last expression the limits $N _A,N_B\gg 1$ and $\epsilon \ll 1$
have been used.
Equation~\eqref{eq:lambdarmt} clearly demonstrates the sensitivity
of a large system to even extraordinarily weak coupling.
Only if $N _A,N_B$, and $\epsilon$ are scaled such that
$\Lambda_{\text{RMT}}$ is kept fixed,
universal statistical behavior of the bipartite system is to be expected.
In contrast, without fixing the scaling of $\Lambda$,
as $N_A\rightarrow \infty$, the transition to effectively random eigenstates
would be discontinuously rapid if $\epsilon>0$.

The evaluation of $\Lambda$ generally is much more complicated in
many-body systems; see~\cite{FreKotPanTom1988b} where it was analyzed for the
compound nucleus in the context of time reversal symmetry breaking.  One
extremely important factor complicating the analysis in most cases is that no
matter how many particles are involved in the system, their interactions are
dominated or limited to one- and two-body operators.  In the standard random
matrix ensembles, if they are being used to model a system of $m$-particles,
the ensemble implicitly is dominated by $m$-body interactions.  If not, there
would have to be a large number of matrix elements put to zero in an
appropriate basis of Slater determinant states.  This fact has led to the study
of the more complicated and difficult to work with so-called embedded
ensembles~\cite{Kot2014}.  In fact, as the number of particles increases,
Hamiltonian matrix elements would become increasingly sparse.  The selection
rules arising through the restriction of the body rank of the interactions thus
create a great deal of matrix element correlations and large scale structure
imposed on local fluctuations.  For the nuclear time reversal breaking
analysis, it was possible to cast the theory as a means to calculate an
effective dimensionality, which is necessarily much smaller than the
dimensionality of the full space.

\subsection{Regularization of perturbation theory}
\label{regularizations}

The main approach to regularize the divergences that arise in the perturbation
theory of symmetry breaking random matrix ensembles was introduced in the
earliest works on the subject~\cite{FreKotPanTom1988,Tom1986}.  There
sometimes arises fluctuation measure specific or symmetry specific aspects, but
the basic approach is common to all cases.  Typically, the leading order is
determined by the regularization required for a pair of levels (eigenphases) to
become nearly degenerate.  The probability of three or more levels is
sufficiently lower as to generate only higher order corrections.

Consider one realization with a pair of levels that are nearly degenerate and
sufficiently isolated.  There is a subset, yet still infinite, series of terms
in the perturbation theory that contains the near vanishing denominators
responsible for the divergence.  Isolating just that series of terms, they have
to be equivalent to the series arising from a two-level system.  Thus,
regularization to lowest order proceeds by using the two level exact solution
to represent the resummation of this infinite series of diverging terms.  The
effective and scaled Hamiltonian, ignoring an irrelevant phase, is
\beq
\left( \begin{array}{cc}
s/2     &\sqrt{\Lambda w} \\
\sqrt{\Lambda w}  &-s/2
 \end{array} \right),
\eeq
where the $jk$ subscript is suppressed on $s$ and $w$.
The unperturbed state $(1, 0)^T$ gets rotated by the interaction
$\Lambda>0$ to $(1, a)^T/\sqrt{1+|a|^2}$, where
\beq
|a|^2 = \dfrac{\left(s-\sqrt{s^2+4 \Lambda w}\right)^2}{4 \Lambda w}
      = \dfrac{4 \Lambda w}{(s+\sqrt{s^2+4 \Lambda w})^2}.
\eeq
The weight $\Lambda w/s^2$ of the mixing
as obtained by perturbation theory, is replaced by $|a|^2/(1+|a|^2)$, i.e.\
with a bit of algebra the regularization is seen to involve the replacement
\beq
\label{eq:regularization}
  \dfrac{\Lambda w}{s^2} \mapsto
      \dfrac{1}{2}\left(1-\dfrac{|s|}{\sqrt{s^2+ 4 \Lambda w}}\right).
\eeq
This correctly accounts for near degenerate cases when $s \rightarrow 0$ and
the states are nearly equally superposed. For larger $s$, in the perturbative
regime that we are interested in, $\Lambda \ll 1$, the mapping is effectively
accounting for a part of the higher order corrections.  It does no harm to use
the regularized expression beyond where it is needed as it only effects higher
order corrections.  Such a regularization is better substantiated and more
accurate than providing simply a lower cut-off to divergent integrals in the
spacing $s$.

Making this substitution in Eq.~\eqref{eq:lambda2-again} gives the
expectation values of the first two Schmidt eigenvalues as
\begin{eqnarray}
\overline{\lambda_1} &\approx& 1 -  \int^\infty_{0}{\rm d}s \int_0^\infty {\rm d}w \left(1-\dfrac{ s}{\sqrt{s^2+ 4 \Lambda w}}\right) e^{-w} \nonumber \\
& = & 1-\sqrt{\pi \Lambda} \label{eq:lambda1-avg} \\
\overline{\lambda_2} &\approx &  \int^\infty_{0}{\rm d}s \int_0^\infty {\rm d}w \left(1-\dfrac{ s}{\sqrt{s^2+ 4 \Lambda w}}\right) e^{-w-2s} \nonumber \\
&= & \sqrt{\pi \Lambda}+ 2 e^{-4 \Lambda } \Lambda  \left(\text{Ei}(4 \Lambda )-\pi  \text{erfi}\left(2 \sqrt{\Lambda }\right)\right)
\label{eq:lambda2-avg-full} \\
&= &\sqrt{\pi \Lambda}+2\Lambda [\gamma+\ln (4\Lambda)]-8 \sqrt{\pi} \Lambda^{3/2}+ ...\ , \label{eq:lambda2-avg-approx}
\end{eqnarray}

where the imaginary error function is $\text{erfi}(x)=-\ui \text{erf}(\ui x)$,
$\text{Ei}(x)$ is the exponential integral~\cite[Eq.~6.2.5]{DLMF}, and
$\gamma=0.57721...$ is Euler's constant.  The last line gives the functional
expansion for the small $\Lambda$ regime, a bit beyond the $\Lambda$-order to
which it is valid.

As mentioned previously, the lowest order behavior is augmented from $\Lambda$
to $\sqrt{\Lambda}$ once the divergence is regularized properly.  It also turns
out that
$\overline{\lambda_1} +\overline{\lambda_2}
  =1+\mathcal{O}(\Lambda \ln \Lambda)$.
The third and smaller Schmidt eigenvalues must contribute in higher orders of
the transition parameter.  This fact is crucial for the development of the
recursively embedded perturbation theory in
Sect.~\ref{sec:recursive-perturbation-theory}.

\subsection{A Floquet system: coupled kicked rotors}
\label{sec:kicked-rotors}

It is instructive to have a simple, bipartite, deterministic, chaotic dynamical
system to compare against the universal transition results
emerging from random matrix theory for the rest of the paper.
Simple in this context means that it is possible
both to calculate the value of $\Lambda$ as a function of interaction strength
analytically, and to carry out calculations for reasonably large values of
$N_A,N_B$.  A system of two coupled kicked rotors quantized on the unit torus
is ideal. Here we will restrict the numerical computations to the case
$N_A=N_B=N$.

A rather general class of interacting bipartite systems can be described by the
unitary Floquet operator of the form given in Eq.~(\ref{eq:Uab}), including
many-body systems.  A specific example, leading to the kicked rotors, for which
the quantum time-evolution operator has this form begins with the Hamiltonian
\begin{equation} \label{eq:hamiltonian-again}
\begin{split}
 H  = & \frac{1}{2}(\pA^2+\pB^2) \\
      & + \bigl[\VA(\qA)+\VB(\qB)+ b \,\VAB (\qA,\qB)\bigr] \,\delta_t
\end{split}
\end{equation}
where $\delta_t= \sum_{n=-\infty}^{\infty} \delta(t-n)$ is a periodic sequence
of kicks with unit time as the kicking period.
The subsystem one-kick unitary propagators connecting states are given by
\begin{equation}
U_A=\exp[-\ui p_A^2/(2 \hbar)] \exp(-\ui V_A/\hbar)
\end{equation}
and likewise for $B$.  The interaction or entangling operator is
\begin{equation}
U_{AB}(b)=\exp(-i b V_{AB}/\hbar).
\end{equation}

The classical system is well defined and given by a $4$-dimensional symplectic
map with the interaction strength tuned by the parameter $b$.  A particularly
important and well studied example are the coupled kicked
rotors~\cite{Fro1971,Fro1972}, which have been realized in experiments on cold
atoms~\cite{GadReeKriSch2013}.  The most elementary case is for two interacting
rotors with the single particle potentials
\begin{equation}
 V_A=K_A \cos(2 \pi q_A)/4 \pi^2
\end{equation}
similarly for $B$, and the coupling interaction
\begin{equation}
 \VAB=\frac{1}{4 \pi^2}\, \cos[2 \pi(\qA+\qB)].
 \label{eq:intpot}
\end{equation}
The unit periodicity in the angle variables $q_j$ is extended here to the
momenta $p_j$ so that the phase space is a 4-dimensional torus.

If the kicking strengths $\KA$ and $K_B$ of the individual maps are each chosen
sufficiently large, the maps in $(\qA, \pA)$ and $(\qB, \pB)$ are strongly
chaotic with a Lyapunov exponent of approximately $\ln (K/2)$~\cite{Chi1979}.
There are some special ranges of $K$ to avoid corresponding to small scale
islands formed by the so-called accelerator modes~\cite{Chi1979}.  This is
supported by recent rigorous results showing that the stochastic sea of the
standard map has full Hausdorff dimension for sufficiently large generic
parameters~\cite{Gor2012}.  In this article $\KA=10$ and $\KB=9$ are used
everywhere in which case the dynamics of the uncoupled maps numerically is
sufficiently chaotic, i.e.\ no regular islands exist on a scale large enough to
matter.

The quantum mechanics on a torus phase space of a single rotor leads to a
finite Hilbert space of dimension $N$, see e.g.\ Refs.~\cite{BerBalTabVor1979,
  HanBer1980, ChaShi1986, KeaMezRob1999, DegGra2003bwcrossref}.
The effective Planck constant is $h=1/N$.  Thus $U_A$ and $U_B$ are
$N$--dimensional unitary operators on their respective spaces.  Explicitly, the
unitary matrix $U_A$ reads
\begin{align}
\label{2dqmap}
U_A(n^{\prime},n)\,=\, \frac{1}{N}\,  \exp\left(- i N
\frac{K_A}{2 \pi} \cos\left(\frac{2\pi}{N}(n+\theta_p)\right)\right) \nonumber \\
\times
\, \sum_{m=0}^{N-1}\exp\left(-\frac{\pi i }{N} (m+\theta_q)^2\right) \,
\exp\left(\frac{2 \pi i}{N} (m+\theta_q)(n-n^{\prime})\right).
\end{align}
where $n,n'\in\{0,1, ..., N-1\}$.
The parameters $\theta_q, \theta_p \in[0, 1[$ allow for shifting
the position grid by which any present symmetries, in particular parity
and time-reversal, can be broken.

If $U_{AB}(b)$ is only a function of position, such as in
Eq.~\eqref{eq:hamiltonian-again}, this can be represented in position space by
a diagonal matrix.  Thus, the resulting $N^2\times N^2$--dimensional unitary
matrix for the coupled kicked rotors is given by
\begin{align}
\label{4dqmap}
\br n_1^{\prime} n_2^{\prime}|{\cal U}_{\text{KR}}|n_1 n_2 \kt \,=\,
U_A(n_1^{\prime},n_1)\,
U_B(n_2^{\prime},n_2)\,\nonumber \\
\times \exp\left( - i N  \frac{b}{2 \pi} \cos(\frac{2 \pi}{N}
(n_1+n_2+2 \theta_p))\right).
\end{align}
The choice $N=100$ is taken as sufficiently large to give asymptotic results
and the phases $\theta_q=0.34$ and $\theta_p=0.24$ are chosen to break
time-reversal and parity symmetry, respectively.  Such coupled quantum maps
have been studied in different contexts~\cite{Lak2001,RicLanBaeKet2014} where
more details can be found.  Note that coupled kicked rotors with a different
interaction term have also been studied on the cylinder~\cite{She1994}.

Applying Eq.~(\ref{eq:lambda1-new}) to the coupled kicked rotors gives
\begin{equation}
\label{eq:lambda-stdmap}
 \Lambda_{\text{KR}} = \frac{N^2}{4\pi^2} \left(1-J_0^2(Nb/2 \pi)\right)
         \approx \frac{N^4 b^2}{32\pi^4},
\end{equation}
where the approximation holds for $Nb \ll 1$.  Using the approximations in
Eqs.~\eqref{eq:lambdarmt} and \eqref{eq:lambda-stdmap} relates the kicked rotor
interaction strength $b$ to the parameter $\varepsilon$ of the random matrix
model explicitly, i.e.
\begin{equation} \label{eq:epb1}
\epsilon = \sqrt{\frac{3}{8\pi^4}} \,  N b,
\end{equation}
for small values of $\epsilon$ and $Nb$, and $N\gg1$.
In practice this approximation is good for the entire
transition even if $N$ is only moderately large.

\begin{figure}[t]
\includegraphics{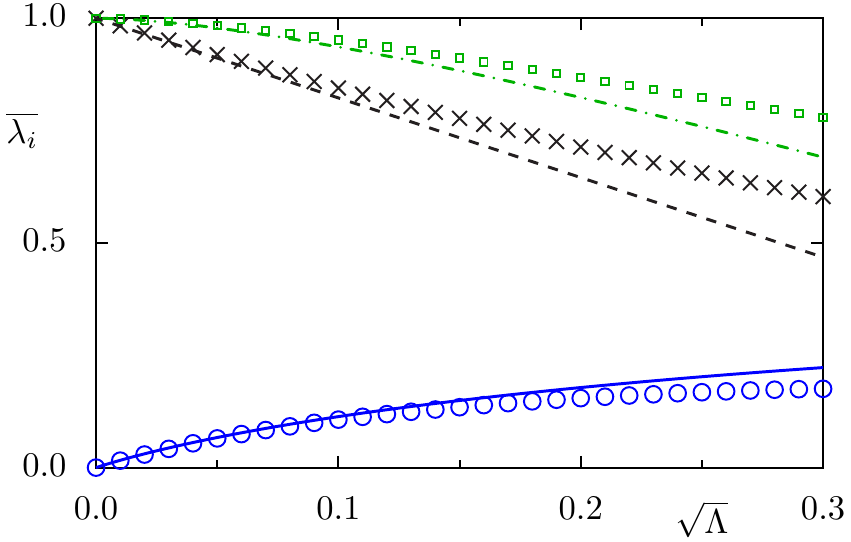}
\caption{\label{fig:averaged-largest-eigenvalues}
Average Schmidt eigenvalues $\overline{\lambda_1}$ (black crosses)
and $\overline{\lambda_2}$ (blue circles)
and their sum (green squares)
for the coupled kicked rotors versus $\sqrt{\Lambda}$.
Comparison with the corresponding predictions
\eqref{eq:lambda1-avg}, dashed black line,
\eqref{eq:lambda2-avg-full}, full blue curve,
and their sum, dashed-dotted green curve.}
\end{figure}

\begin{figure}[t]
\includegraphics{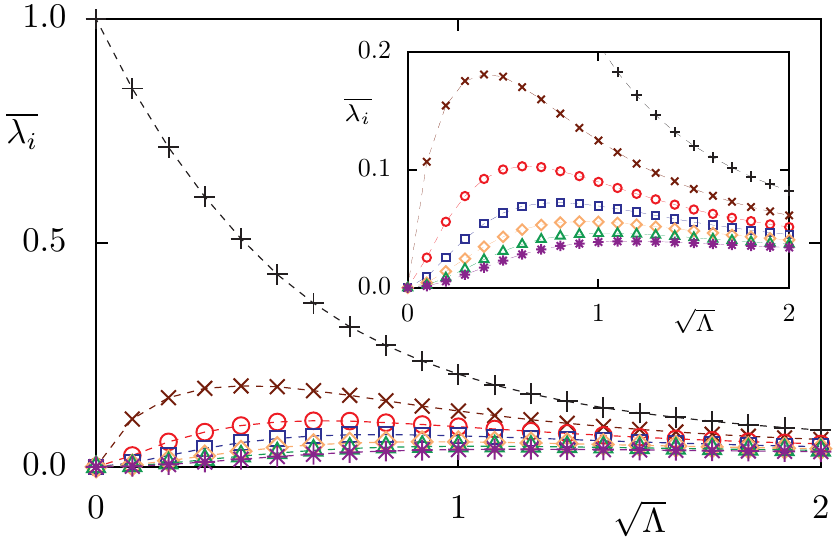}
\caption{\label{fig:lambda_i-vs-Lambda}
Average Schmidt eigenvalues $\overline{\lambda_i}$, $i=1, 2, ..., 7$,
(symbols with lines as guide to the eye, from top to bottom)
for the coupled kicked rotors versus $\sqrt{\Lambda}$.}
\end{figure}

Let us first compare the first two mean Schmidt eigenvalues of the coupled
kicked rotors to the random matrix ensemble predictions, see
Fig.~\ref{fig:averaged-largest-eigenvalues}.
The averaging is a spectral
average over all $N^2$ eigenstates for the coupled kicked rotors.
The perturbative nature of the
expression Eq.~\eqref{eq:lambda1-avg} for $\overline{\lambda_1}$
is evident showing that the derivation
only provides the initial dependence at $\Lambda=0$ as it is designed to do.
For $\overline{\lambda_2}$ the validity of the result
\eqref{eq:lambda2-avg-full} extends beyond the initial dependence as given by
Eq.~\eqref{eq:lambda2-avg-approx}.

\section{Eigenvalue moments of the reduced density matrix}
\label{momentsrdm}

For a global view of the spectrum of the reduced density matrix as the coupling
is increased, Fig.~\ref{fig:lambda_i-vs-Lambda} shows the behavior of various
$\overline{\lambda_i}$ versus $\sqrt{\Lambda}$.
Here, and in the following we restrict to $N_A = N_B = N$
with $N=100$ chosen for the numerical computations.
It is seen
from the figure that whereas $\overline{\lambda_1}$ monotonically decreases
towards its random matrix average of $\approx 4/N$,
other principal eigenvalues, such as
the second or third largest, display a non-monotonic approach to their
respective averages. The {\it smallest} eigenvalue, $\overline{\lambda_N}$
grows to about $1/N^3$, which is the
random matrix average~\cite{Zni2007,MajBohLak2008}.
For $\Lambda \gg 1$, the probability density of the set of $\lambda_i $
follows the
\Marcenko-Pastur distribution given in Eq.~\eqref{eq:marcenko-pastur-law}.
The largest alone is distributed according to the Tracy-Widom density (after
appropriate scaling and shift)~\cite{TraWid1996}, whereas the smallest
is known to be exponential~\cite{MajBohLak2008}.
Numerical results for the transition towards these results are presented in
Sect.~\ref{sec:distribution-of-schmidt-eigenvalues}.  Also see the recent
paper~\cite{KumSamAna2017} for additional random matrix results concerning
the smallest eigenvalue and applications to coupled kicked tops.

\subsection{General moments}
\label{generalmoments}

To characterize the entanglement in bipartite systems the general moments
$\mu_{\alpha}$ of Eq.~(\ref{eq:momentsdefn}) and thus the sum of the averages
$\overline{\lambda_j^{\alpha}}$ are needed.  The relationship of the entropies
to the general moments is given in Eq.~\eqref{eq:Tsallis}.  In the perturbative
regime, the largest eigenvalue $\lambda_1$ decreases from 1, see
Fig.~\ref{fig:averaged-largest-eigenvalues}, and thus has a different behavior
compared to all other eigenvalues.  This is clear from the content of
Eqs.~\eqref{eq:lambda1} and \eqref{eq:lambda2}
where each eigenvalue $\lambda_j$ with $j>1$ has its own expression,
while the largest follows from normalization,
$\lambda_1=1-\sum_{j>1} \lambda_j$.
Therefore it is necessary to treat the general
moments of $\lambda_1$ separately from the others as
it requires some additional analysis.

\subsubsection{Moments $\sum_{j>1}\lambda_j^{\alpha}$}

The moment expressions for the $\sum_{j>1} \lambda_j^{\alpha}$ can be written down
immediately from the results of Sects.~\ref{erdmea},
\ref{regularizations}. There is no need to invoke the closer next neighbor
statistics as there is a sum over all the eigenvalues $j>1$ of the unperturbed
Hamiltonian and hence $R_2(s)=1$.  This gives for the $\alpha$ ($ > 1/2$)
moment
\begin{eqnarray}
\label{highjmoment}
&&\sum_{j>1}\overline{\lambda_j^\alpha} = \nonumber \\
&& \int^\infty_{-\infty}{\rm d}s \int_0^\infty \frac{{\rm d}w}{2^\alpha}\left(1-\dfrac{ |s|}{\sqrt{s^2+ 4 \Lambda w}}\right)^\alpha \overline{R(s, w)} \nonumber \\
&=& \int^\infty_{-\infty}{\rm d}s \int_0^\infty \frac{{\rm d}w}{2^\alpha}\left(1-\dfrac{ |s|}{\sqrt{s^2+ 4 \Lambda w}}\right)^\alpha e^{-w} \nonumber \\
&= & C_2(\alpha)\sqrt{\Lambda}
\end{eqnarray}
where
\beq
\begin{split}
C_2(\alpha)
&=2\sqrt{\pi} \int_0^{\pi/2}\dfrac{1}{\sin^2\theta}
  \sin^{2 \alpha}\frac{\theta}{2}\, d \theta\\
&=\dfrac{\sqrt{\pi}}{2} \int_0^{1/2}\dfrac{t^{\alpha}}{t^{3/2} (1-t)^{3/2}} dt\\
&= \dfrac{\sqrt{\pi}}{2} \text{B}_{1/2} \left(\alpha-\frac{1}{2},-\frac{1}{2} \right), \;\; \alpha>1/2.
\end{split}
\label{eq:C-2-alpha}
\eeq
Here $B_z(a,b)$ is the incomplete Beta function \cite[Eq.~8.17.1]{DLMF}
defined as
\begin{equation}
\begin{split}
 B_z(a,b) & = \int_0^z t^{a-1} (1-t)^{b-1} \, dt\\
          & = \frac{z^a}{a} \; _2 F_1\left(a, 1-b; a+1;x\right).
\end{split}
\end{equation}

Note that the evaluation of the integral is exact and that all moments
($\alpha > 1/2$) are proportional to $\sqrt{\Lambda}$, i.e.\
there are no higher order terms coming from these expressions.
If $\alpha = 1/2$ there is a divergence ($C_2(1/2)=\infty$) and this is a
special value as far as the moments are concerned. This is due to the
contributions of the very small eigenvalues.  It turns out that for
$\alpha=1/2$ the order is no longer $\sqrt{\Lambda}$.  This will be dealt with
later in Sect.~\ref{sec:moment-half}.

The result for $\alpha=1$, which follows from $C_2(1)=\sqrt{\pi}$, is identical
to the leading order of the average second largest eigenvalue in
Eq.~(\ref{eq:lambda2-avg-approx}).  The reason for this is that the other
eigenvalues do not contribute to this order. This really justifies the use of
only {\it two} reduced density matrix eigenvalues in the perturbative regime.
Another special case corresponding to $\alpha=2$ gives
\beq \label{eq:sum-for-alpha-2}
\sum_{j>1} \overline{\lambda_j^{2} } = \sqrt{\pi \Lambda}\,(1-\pi/4).
\eeq

\subsubsection{Moments of $\lambda_1$}

Turning to the largest eigenvalue, it is given by
\begin{eqnarray}
&&\overline{\lambda_1^{\alpha}} = \overline{\left(1-\sum_{j>1}\lambda_j\right)^{\alpha}} \nonumber \\
&& = \overline{\left(1 -  \int^\infty_{-\infty}{\rm d}s \int_0^\infty \frac{{\rm d}w}{2} \left(1-\dfrac{ |s|}{\sqrt{s^2+ 4 \Lambda w}}\right) R_2(s,w)\right)^\alpha}. \nonumber \\
\end{eqnarray}
The extra ingredient not present for the moments of the other eigenvalues is
the ensemble averaging of the powers of the density, $R(s, w)$, are also
involved.  To see how this changes the moment calculations, consider the
simplest case, the quadratic terms in the binomial expansions of the moments.
There is a quadruple integral for which one needs the ensemble average of
$R(s_1, w_1)R(s_2, w_2)$.  It has two contributions, those coming from
off-diagonal terms in the products of $\delta$-functions from
Eq.~\eqref{twopointws} and the diagonal terms.  Thus,
\begin{eqnarray}
&& \overline{R(s, w)R(s^\prime, w^\prime)} \nonumber \\
&=& \rho(w) \rho(w^\prime)R_3(s,s^\prime)
+ \delta\left(w-w^\prime\right)\rho(w) \delta\left(s-s^\prime\right)R_2(s) \nonumber \\
&=& e^{-w-w^\prime} + \delta\left(s-s^\prime\right)\delta\left(w-w^\prime\right) e^{-w}.
\end{eqnarray}
The second line follows due to the independence of the matrix elements from
each other and the spectrum, and the fact that it is the three-point
correlation function of the spectrum that enters.  Fortunately, for Poisson
sequences, for all $k$, $R_k=1$, and that gives the last line.

The second moment therefore is given by
\begin{eqnarray}
\label{secondmoment}
\overline{\lambda_1^2} &=& 1-2 \int^\infty_{-\infty}{\rm d}s \int_0^\infty \frac{{\rm d}w}{2} \left(1-\dfrac{ |s|}{\sqrt{s^2+ 4 \Lambda w}}\right) e^{-w} + \nonumber \\
&& \int^\infty_{-\infty}{\rm d}s \int_0^\infty \frac{{\rm d}w}{4} \left(1-\dfrac{ |s|}{\sqrt{s^2+ 4 \Lambda w}}\right)^2 e^{-w} +\nonumber \\
&& \left[\int^\infty_{-\infty}{\rm d}s \int_0^\infty \frac{{\rm d}w}{2} \left(1-\dfrac{ |s|}{\sqrt{s^2+ 4 \Lambda w}}\right)e^{-w}\right]^2 \nonumber \\
&=& 1-2C_2(1)\sqrt{\Lambda} +C_2(2) \sqrt{\Lambda} +C_2^2(1)\Lambda \nonumber \\
&=& 1-\left(1+\frac{\pi}{4}\right)\sqrt{\pi\Lambda} +\pi\Lambda.
\end{eqnarray}
It is immediately apparent that the leading order terms proportional to
$\sqrt{\Lambda}$ come from the diagonal terms for which all the energy and
matrix element variables are reduced to the mimimum set.  Note that there is
only one way for all the variables to be maximally correlated; i.e.\
the sets of integration variables $\{w_j\},\{s_j\}$ reduce to $(w,s)$.
There are
corrections, depending on the moment $\alpha$ considered, polynomial in
$\sqrt{\Lambda}$.  The next to leading order, $\mathcal{O}(\Lambda)$, come from
terms whose integrals can be reduced to $(w_1,w_2,s_1,s_2)$.  In the second
moment example shown above, there is only one term that is of this form.
However, for arbitrary moments, there are a variety of combinations of
possibilities that form a sub-binomial expansion given ahead.

\subsubsection{The leading order of $\lambda_1$ moments}

Beginning with the binomial expansion of the moments, the leading order comes
from the terms remaining after reducing the power of summations to a single
summation
\begin{eqnarray}
\overline{\lambda_1^{\alpha}} &=& \overline{\left(1-\sum_{j>1}\lambda_j\right)^{\alpha}} \nonumber \\
&=& \overline{\left(\sum_{k=0}^\infty  -1^k\left(\alpha \atop k \right)\left[\sum_{j>1}\lambda_j \right]^k \right)} \nonumber \\
& \mapsto & \overline{\left(\sum_{k=0}^\infty  -1^k\left(\alpha \atop k \right)\sum_{j>1}\lambda_j^k \right)} .
\end{eqnarray}
By inverting the order of the remaining summations, the series can be resummed
to give a compact expression for arbitrary moments.  The zero$^{th}$ order
terms have to be handled separately.  This gives,
\begin{eqnarray}
\label{eq:ensemble-averaged-moments}
\overline{\lambda_1^{\alpha}} &=& \overline{1 + \sum_{j>1} \sum_{k=1}^\infty  -1^k\left(\alpha \atop k \right) \lambda_j^k} \nonumber \\
&=& \overline{1 + \sum_{j>1}\left[\left(1-\lambda_j\right)^\alpha -1\right]} \nonumber \\
&=& 1-C_1(\alpha)  \sqrt{\Lambda},
\end{eqnarray}
where
\beq
\label{c1a}
\begin{split}
C_1(\alpha)
 &=  2\sqrt{\pi} \int_0^{\pi/2} \frac{1}{\sin^2\theta}
               \left( 1-\cos^{2 \alpha}\frac{\theta}{2}\right) \, d \theta\\
 &= \frac{\sqrt{\pi}}{2}\int_0^{1/2}\frac{1-(1-t)^{\alpha}}{t^{3/2} (1-t)^{3/2}} dt\\
 &= \sqrt{2 \pi} \; _2F_1 \left(-\frac{1}{2},\frac{3}{2}-\alpha;\frac{1}{2};\frac{1}{2} \right).
\end{split}
\eeq
Here $_2F_1(\cdot)$ is the Gauss hypergeometric function
\cite[Eq.~15.2.1]{DLMF} defined as
\[
   _2F_1 \left(a,b;c;z\right)
   =1+\frac{a\, b}{1!\,c}z +\frac{a(a+1)\, b(b+1)}{2!\, c(c+1) }z^2+\cdots.
\]
With the analytic results for $C_1(\alpha)$ and $C_2(\alpha)$, the results of
Eq.~(\ref{eq:lambda2-avg-approx}) have been generalized to complete general
moments.  $C_1(1)=\sqrt{\pi}$ is consistent with the expression given there,
and $C_1(2)=1+\tfrac{\pi}{4}$
is consistent with Eq.~\eqref{secondmoment}.  A moment of special interest
ahead is
\beq
\overline{\sqrt{\lambda_1}} = 1-\sqrt{\pi \Lambda}\, [\sqrt{2}-\ln(1+\sqrt{2})],
\label{eq:lamb1momhalf}
\eeq
due to its relationship to the nearest maximally entangled state.

\subsubsection{The $\mathcal{O}(\Lambda)$ correction of the $\lambda_1$ moments}

\begin{figure}
\begin{center}
\includegraphics{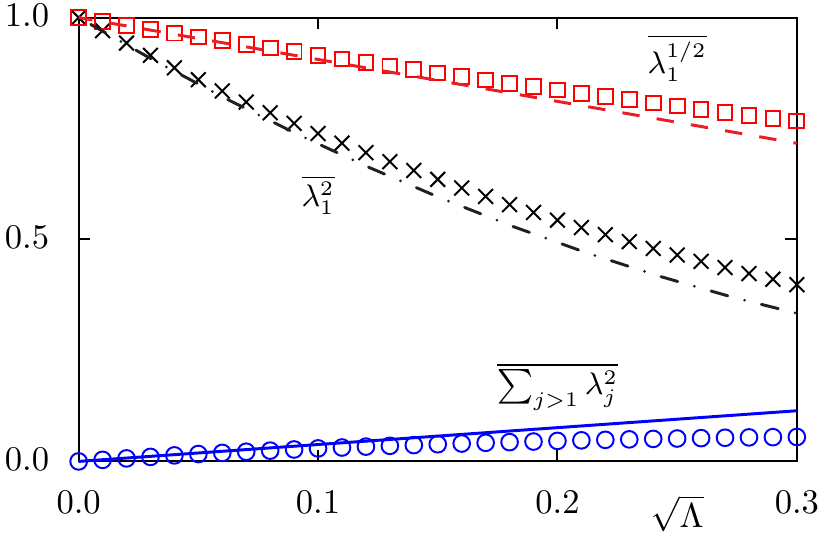}
\end{center}
\caption{\label{fig:avg-lambda-moments}
Average moments $\overline{\lambda_1^2}$ (black crosses),
$\overline{\lambda_1^{1/2}}$ (red squares), and
$\sum_{j>1}\overline{\lambda_j^{2}}$ (blue circles)
for the coupled kicked rotors versus $\sqrt{\Lambda}$.
These are compared to their corresponding perturbative results,
Eq.~\eqref{secondmoment}, dashed-dotted line,
Eq.~\eqref{eq:lamb1momhalf}, dashed red line,
and Eq.~\eqref{eq:sum-for-alpha-2}, full blue line, respectively.}
\end{figure}

In order to calculate the $\mathcal{O}(\Lambda)$ correction of the $\lambda_1$
moments, the key question is the combinatoric one of separating the eigenvalues
in each term of $\left[\sum_{j>1}\lambda_j \right]^k$ into two groups; of
course, $k\ge2$.  The results is
\begin{eqnarray}
\left[\sum_{j>1}\lambda_j \right]^k &\mapsto& \frac{1}{2}\sum_{l=1}^{k-1}\left( k \atop l\right)  \sum_{j>1} \sum_{j^\prime>1\ne j} \lambda_j^l \lambda_{j^\prime}^{k-l} \nonumber \\
&=& \frac{1}{2} \sum_{j>1} \sum_{j^\prime>1\ne j}\sum_{l=1}^{k-1}\left( k \atop l\right)  \lambda_j^l \lambda_{j^\prime}^{k-l} \nonumber \\
&=&  \frac{1}{2} \sum_{j>1} \sum_{j^\prime>1\ne j} \left[\left(\lambda_j+\lambda_{j^\prime}\right)^k - \lambda_j^k-\lambda_{j^\prime}^k\right]. \nonumber \\
\end{eqnarray}
Therefore, the $\mathcal{O}(\Lambda)$
correction $C_3(\alpha)\Lambda$ is given by
\begin{eqnarray}
&&C_3(\alpha)\Lambda = \nonumber \\
&&\frac{1}{2} \sum_{j>1} \sum_{j^\prime>1\ne j}  \overline{\left(\sum_{k=2}^\infty  -1^k\left(\alpha \atop k \right)\left[\left(\lambda_j+\lambda_{j^\prime}\right)^k - \lambda_j^k-\lambda_{j^\prime}^k\right] \right)} \nonumber \\
&&=\frac{1}{2} \sum_{j>1} \sum_{{j^\prime>1} \atop \ne j}  \overline{1+\left(1-\lambda_j-\lambda_{j^\prime}\right)^\alpha - \left(1-\lambda_j\right)^\alpha-\left(1-\lambda_{j^\prime}\right)^\alpha }, \nonumber \\
\end{eqnarray}
which after using the same manipulations as for the earlier integrals leads to
\begin{eqnarray}
&&C_3(\alpha) = \frac{\pi}{8} \int_0^{1/2} {\rm d}t_1 \int_0^{1/2} {\rm d}t_2 \times \nonumber \\
&& \frac{\left[1+(1-t_1-t_2)^\alpha - (1-t_1)^\alpha- (1-t_2)^\alpha\right]}{t_1^{3/2} (1-t_1)^{3/2} t_2^{3/2} (1-t_2)^{3/2}}.
\end{eqnarray}
Note that $C_3(2)=\pi$, which is consistent with Eq.~\eqref{secondmoment}.
For some other values of $\alpha$ relevant in the following,
the integral evaluates to $C_3(3)=\frac{3 \pi ^2}{4}$ and
$C_3(4)=2 \pi +\frac{3 \pi ^3}{16}$.
The computations of $\overline{\lambda_1^2}$, $\overline{\lambda_1^{1/2}}$, and
$ \sum_{j>1}\overline{\lambda_j^{2}}$ for the coupled kicked rotors are shown
in Fig.~\ref{fig:avg-lambda-moments} and compared to the analytic results of
Eqs.~\eqref{secondmoment} and \eqref{eq:lamb1momhalf}, and
Eq.~\eqref{eq:sum-for-alpha-2}, respectively.  The agreement in the
perturbative regime is quite good.

\subsection{Entropies}

With these results for the average eigenvalue moments of the reduced density
matrix, everything needed for the entropies defined in
Eq.~\eqref{eq:momentsdefn} has been evaluated.
Combining Eq.~\eqref{eq:sum-for-alpha-2} with Eq.~(\ref{secondmoment})
results in
\beq
\label{eq:moments2}
\overline{\mu_2}= \sum_{j=1}^N \overline{\lambda_j^{2}}
     = \overline{\tr({\rhoA}^2)}
     = 1-\frac{\pi^{3/2}}{2} \sqrt{\Lambda} + \pi\Lambda,
\eeq
where the neglected terms are presumably of $\mathcal{O}(\Lambda^{3/2})$. This
is the random matrix prediction for the purity of the density matrix of generic
eigenstates of chaotic subsystems that are perturbatively entangled due to the
coupling.  As decoherence, i.e.\ coupling of a system to the environment is
usually the cause of loss of purity, this shows the universal manner in which
``decoherence" due to coupling with a chaotic subsystem results in the
degradation of the purity of eigenstates.

The generalized moments for $\alpha> 1/2$ are
\beq
\label{eq:momentsalph}
\overline{\mu_{\alpha}} = \sum_{j=1}^N \overline{\lambda_j^{\alpha}}
          = 1-C(\alpha) \sqrt{\Lambda} + C_3(\alpha) \Lambda + ...
\eeq
where
\beq
\label{eq:Calpha}
\begin{split}
C(\alpha)
&=C_1(\alpha)-C_2(\alpha)
 = \frac{\sqrt{\pi}}{4}\int_0^1 \dfrac{1-t^{\alpha}-(1-t)^{\alpha}}{t^{3/2} (1-t)^{3/2}} dt\\
&=\pi \frac{\Gamma(\alpha-\nicefrac{1}{2})}{\Gamma(\alpha-1)}.
\end{split}
\eeq
The incomplete Beta functions of $C_1(\alpha)$ and $C_2(\alpha)$ combine to
produce complete Beta functions.  This perturbative moment evaluation is one of
the central results of this work.  Note that $C(1)=C_3(0)=0$, correctly
reproducing the unit trace of the density matrix, and that $C(\alpha)<0$ for
$\alpha<1$.  In the regime when $\alpha<1$ the smaller eigenvalues start to
become more important as well.  The critical value $\alpha=1$ corresponds to
the von Neumann entropy and is of central interest in quantum information.

Thus it follows that the average HCT entropies,
defined in Eq.~(\ref{eq:Tsallis}), for small $\Lambda$ are
\beq
\label{eq:hct1}
\overline{S_{\alpha}}=\pi \frac{\Gamma(\alpha-1/2)}{\Gamma(\alpha)} \sqrt{\Lambda}-\frac{C_3(\alpha)}{\alpha-1}\Lambda.
\eeq
and the $\lim_{\alpha\rightarrow 1}C_3(\alpha)/(\alpha-1)=C_{3\text{L}}(1)$
is the integral
\begin{eqnarray}
C_{3\text{L}}(1) &&= \frac{\pi}{8} \int_0^{1/2}  \int_0^{1/2}  \frac{{\rm d}t_1 {\rm d}t_2 }{t_1^{3/2} (1-t_1)^{3/2} t_2^{3/2} (1-t_2)^{3/2}} \times \nonumber \\
&& \qquad \left[(1-t_1-t_2)\ln(1-t_1-t_2) \right.\nonumber\\
&& \qquad \left.  - (1-t_1)\ln (1-t_1)- (1-t_2)\ln(1-t_2)\right] \nonumber \\
&&= \frac{\pi^2}{4}(4-\pi).
\end{eqnarray}
Hence the von Neumann entropy, the measure of entanglement in
bipartite pure states, is perturbatively
\beq
\label{eq:S-1-perturb}
\overline{S_{1}} = \pi^{3/2}\sqrt{\Lambda}-C_{3\text{L}}(1)\Lambda.
\eeq

\begin{figure}
\begin{center}
\includegraphics{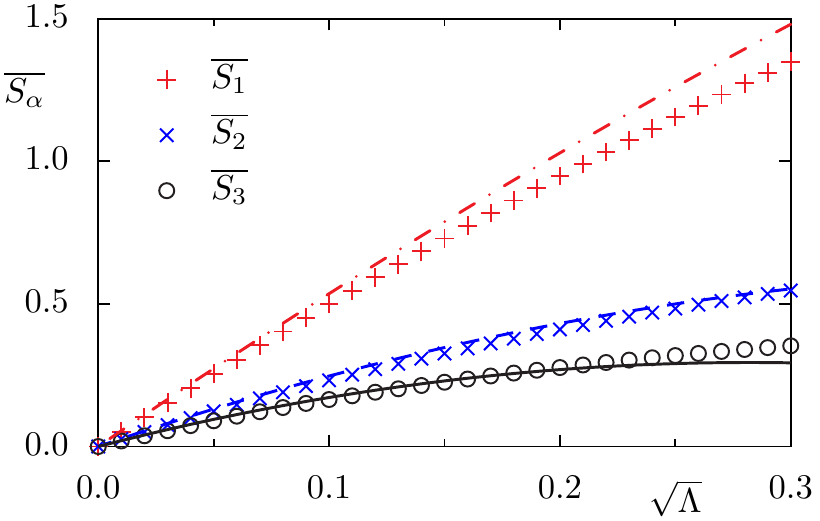}
\end{center}
\caption{\label{fig:avg-entropies}%
Average entropies $\overline{S_\alpha}$ for $\alpha=1, 2, 3$
for the coupled kicked rotors (symbols) in comparison with the
perturbative results, Eq.~\eqref{eq:hct1}, full line
for $\overline{S_3}$ and dashed line for $\overline{S_2}$,
and Eq.~\eqref{eq:S-1-perturb} for $\overline{S_1}$, dashed-dotted line.
}
\end{figure}

Figure~\ref{fig:avg-entropies} shows the average entropies computed for the
coupled kicked rotors in comparison with Eq.~\eqref{eq:hct1}.  Again it is
clear that the numerics for the coupled kicked rotors supports these
perturbative results, but that for larger $\Lambda$ there are
deviations that grow with the coupling.
Also, the von Neumann entropy grows at a faster rate perturbatively
($\Lambda \ll 1$) than the other entropies, in particular the linear
entropy. On the other hand, later it is seen that the von Neumann entropy
approaches its asymptotic ($\Lambda \gg 1$) random matrix value,
the slowest among the shown entropies, including the linear.

\subsection{Moment at $\alpha=1/2$ and distance to the closest maximally entangled state.}
\label{sec:moment-half}

The averaged moment for $\alpha=1/2$ indicates the distance of the
eigenfunction to the closest maximally entangled state.  In the quantum
information context the ``singlet fraction"~\cite{HorHorHor1999} essentially
measures the same quantity.  It is the highest $\alpha$ value for which the
moment depends on subsystem size; note the trivial case of $\alpha=0$
for which the moment is simply the subsystem dimension $N$.
The case $\alpha=1/2$ is marginal and the moment is shown to grow
as a logarithm of system dimensionality,
whereas for $\alpha>1/2$ the moments are independent (except through
the definition of $\Lambda$).  This signals a breakdown of the description
with a single universal dimensionless parameter, and the smaller Schmidt
eigenvalues all contribute significantly to the average value of the moments.

As in the case of regularization, the divergence in Eq.~\eqref{eq:hct1} for
$\alpha=1/2$ is indicative of a change in the functional $\Lambda$ dependency.
However the largest eigenvalue moment $\overline{\lambda_1^{1/2}}$ given in
Eq.~(\ref{eq:lamb1momhalf}), even with the $C_3(1/2)\Lambda$ correction is
still valid.  The critical quantity is the integral in Eq.~(\ref{highjmoment}).
As the value of $\alpha$ decreases, the importance of distant levels
increases.  Although, the number of Schmidt eigenvalues is equal to $N$, for
$\alpha > 1/2$ the decay of the integrand is fast enough that in the
$N\rightarrow \infty$ limit there is no difference whether the upper
integration bound is finite or not.  For $\alpha=1/2$ this is not true and the
fact that $N$, no matter how large, has a finite value must be accounted for.
A good approximation is to change the limits of integration over $s$ from
$-\infty \le s\le \infty$ to $-(N-1)/2 \le s\le (N-1)/2$.  With the
substitution $\sqrt{4 \Lambda w} = s\, \tan \theta$ and using the fact that
$4\sqrt{ \Lambda w}/(N-1) \ll1$, the integral in Eq.~(\ref{highjmoment}) leads
the following modification of Eq.~\eqref{c1a},
\beq
C_2(1/2) =  2\sqrt{\pi} \int_{4 \frac{\sqrt{\Lambda w}}{N-1}}^{\pi/2}
      \dfrac{1}{\sin^2\theta} \sin\frac{\theta}{2}\, d \theta.
\eeq
The $\theta$ integral can be done exactly, and again using the fact that $N$
is large and $\Lambda$ small, the following approximation can be derived
\beq
\begin{split}
 \sum_{j>1} \overline{\sqrt{\lambda_j}}
  = \sqrt{\pi \Lambda}\,
      \biggl[ & \ln\left(\frac{2 (N-1)}{\sqrt{\Lambda}}\right)
                 +\frac{\gamma}{2}+\sqrt{2} \\
              & -2-\ln(1+\sqrt{2}) \biggr].
\end{split}
\eeq
Combining this with the result for the $\alpha=1/2$ moment
of the largest eigenvalue in
Eq.~(\ref{eq:lamb1momhalf}) gives the leading order term
\beq
\label{eq:momhalf}
\overline{\mu_{1/2}}
  = 1+\sqrt{\pi \Lambda} \left[ \ln\left(\frac{2 (N-1)}{\sqrt{\Lambda}}\right)
    +\frac{\gamma}{2}-2\right].
\eeq
Thus the diverging $C_2(\alpha)$ as $\alpha \rightarrow 1/2$ gives rise to the
term proportional to $\ln(N/\sqrt{\Lambda})$.  As a consequence we obtain a
dependence on $N$ and as well as a different leading order $\Lambda$ dependency
proportional to $-\sqrt{\Lambda} \ln (\Lambda)$, rather than $\sqrt{\Lambda}$
which is valid for moments $\alpha>1/2$.

\begin{figure}[h]
\includegraphics{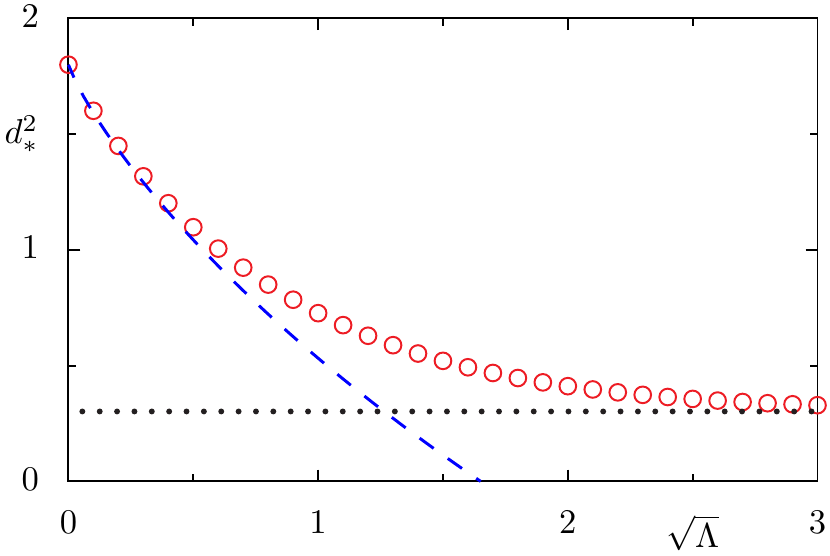}
\caption{\label{fig:dstar} Square of the distance to the closest
maximally entangled state $d_*^2$  (circles) as defined in
Eq.~\eqref{eq:closeststate2} for the coupled kicked rotors as function of
$\sqrt{\Lambda}$.  The dashed line is the perturbative prediction
for $\overline{d_*^2}$ using Eq.~\eqref{eq:distsqavg} and
$(d_*^{\text{product}})^2=1.8$, see Eq.~\eqref{eq:dstar-unentangled-product}.
For larger $\sqrt{\Lambda}$ an approach
towards $\overline{d_*^2}_{\text{RMT}} = 2(1-8/3\pi)$,
see Eq.~\eqref{eq:dstar-random}, (dotted line)
for a typical random state takes place.}
\end{figure}

Not only is the moment $\alpha=1/2$ therefore an interesting limiting moment,
but it is also of importance due to its relationship via
Eq.~(\ref{eq:closeststate2}) to the distance from the closest maximally
entangled state.  It follows from Eq.~\eqref{eq:closeststate2} and
Eq.~\eqref{eq:momhalf} that
\beq
\label{eq:distsqavg}
\overline{d_*^2} = (d_*^{\text{product}})^2- 2\sqrt{\frac{\pi \Lambda}{N}}\, \left[ \ln\left(\frac{2 (N-1)}{\sqrt{\Lambda}}\right) + \frac{\gamma}{2}-2\right].
\eeq
Note that for $\Lambda=0$, the distance is that of a product state as given in
Eq.~(\ref{eq:dstar-unentangled-product}).  Figure~\ref{fig:dstar} shows the
transition of $\overline{d_*^2}$ going from the situation of product states at
$\Lambda=0$ to typical random states at $\sqrt{\Lambda}=3$ for the coupled
kicked rotors.  Eq.~\eqref{eq:distsqavg} describes the initial behavior up to
approximately $\sqrt{\Lambda}=0.5$ very well.

\section{Probability densities of Schmidt eigenvalues and entropies}
\label{sec:distribution-of-schmidt-eigenvalues}

Having completed the perturbation theory of the eigenvalue moments and
entanglement entropies, consider the probability densities of the eigenvalues
of the reduced density matrices.  For $\Lambda \gg1$, the \Marcenko-Pastur law,
Eq.~\eqref{eq:marcenko-pastur-law}, holds for the eigenvalue density of states.
For $\Lambda=0$, the density is a unit $\delta$ function at unity and an
$N-1$ weighted $\delta$ function at the origin.  For very weak interactions,
the density breaks away from the $\delta$ function form limit and is dominated
by the largest and second largest eigenvalues.  Note that the probability
density of $\lambda_1$ in the strongly coupled regime is the extreme value
statistics of Tracy-Widom, as it follows the same universality class of fixed
trace Wishart ensembles~\cite{Nec2007}.

\begin{figure}
\includegraphics{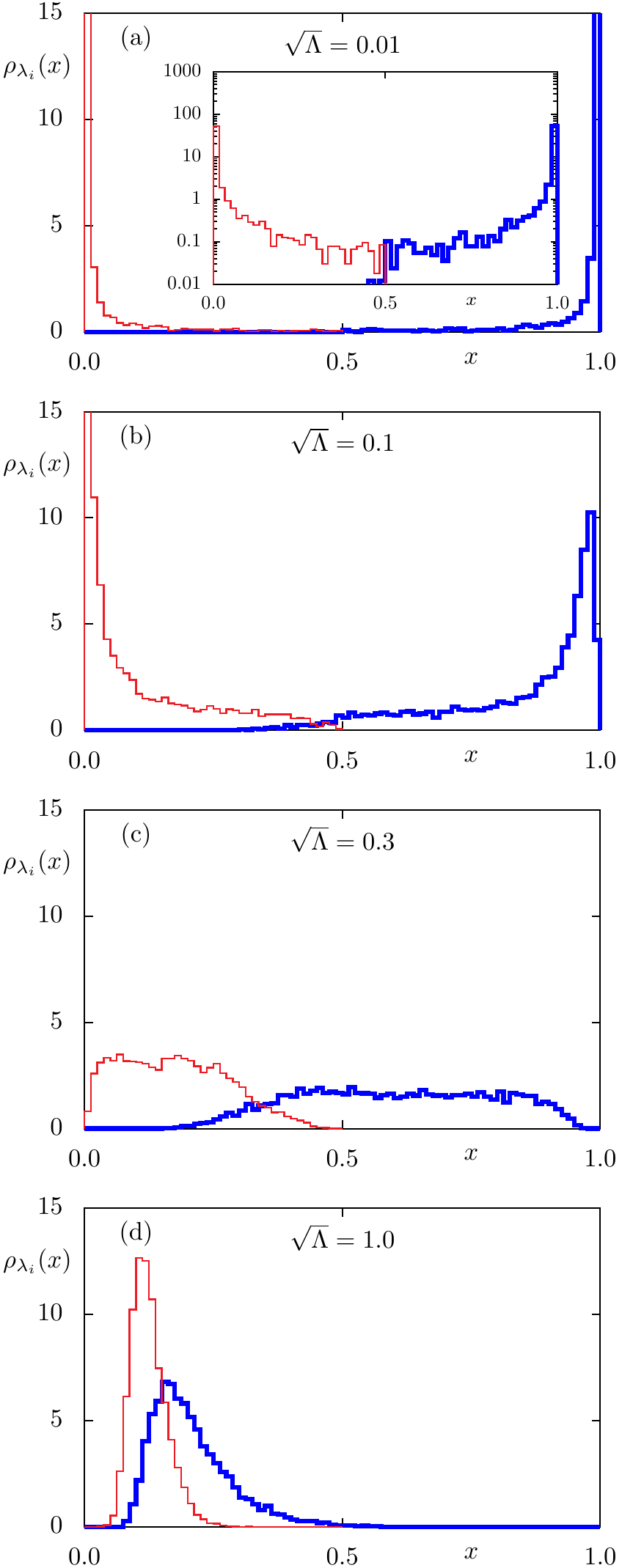}
\caption{\label{fig:distrib-lambda-1-2}
Probability densities $\rho_{\lambda_1}(x)$, thick blue histogram,
and $\rho_{\lambda_2}(x)$, thin red histogram,
for the coupled kicked rotors for
$\sqrt{\Lambda}=0.01$, $0.1$, $0.3$, and $1.0$.}
\end{figure}

Figure~\ref{fig:distrib-lambda-1-2} shows the probability densities
$\rho_{\lambda_1}(x)$ and $\rho_{\lambda_2}(x)$.  For $\Lambda \ll 1$, the
density of the largest eigenvalue $\lambda_1$ is sharply peaked around its
unperturbed value of $1$, while $\lambda_2$ is similarly peaked around $0$. The
two densities appear almost mirror symmetric about $1/2$, and both have
prominent tails extending to $1/2$, indicating instances when both of the them
have large excursions away from their unperturbed values due to near
degeneracies.  A more detailed view of these tails is shortly developed, where
power laws and stable densities exist, and the mirror symmetry is seen to
be an illusion. For moderately larger couplings this picture gets modified.
The probability of the largest eigenvalue develops a tail that crosses $1/2$
and the densities have an overlapping range.  A curious feature that appears
for $\sqrt{\Lambda} \sim 0.3$ is that the largest eigenvalue density is
characterized by an almost uniform density over a wide interval. For even
larger coupling the densities tend to significantly overlap and approach
their random matrix extreme value statistical laws.
Further results in this strong
coupling regime are postponed for discussion in
Sect.~\ref{sec:recursive-perturbation-theory} after the perturbative regime is
considered in detail.

\subsection{Perturbative regime}

Recall that perturbatively the two largest eigenvalues $\lambda_1$ and
$\lambda_2$ are fluctuating variables which from
Eqs.~\eqref{eq:lambda1} and \eqref{eq:lambda2},
after regularization are given by
\beq
\label{eq:lambda12}
\begin{split}
\lambda_1 & =1-\frac{1}{2} \sum_{j}\left(1-\frac{1}{\sqrt{1+4 \Lambda w_j/s_j^2}}\right), \,\\
\lambda_2 & =\frac{1}{2} \left(1-\frac{1}{\sqrt{1+4 \Lambda w/s^2}}\right).
\end{split}
\eeq
where the transition strength $w_j$ and $w$ is distributed according to
$\rho_w(x) = \exp(-x)$. The spacings $s_j$ refer to transtions from a fixed
state to all others; we will take it as independent and uniform in
$[-M/2,M/2]$, where $M \gg 1 $ is the number of terms in the sum for
$\lambda_1$. This leads to a Poisson process for the ordered spacings. The
spacing $s$ in $\lambda_2$ is the \CNtxt{} used in Eq.~(\ref{eq:lambda2-again})
and is distributed according $\rho_\CN(s) = 2 \exp(-2s)$.

\subsubsection{Probability density of the second largest eigenvalue}

Treating $\lambda_2$ first, the second part of Eq.~(\ref{eq:lambda12})
and the subsequent considerations imply that its probability density
is given by
\beq
\begin{split}
\label{eq:distlamb2-1}
&\rho_{\lambda_2}(x)= \\
&\int_{0}^{\infty} 2 e^{-2s}ds \int_0^{\infty}e^{-w} dw\   \delta \left[ x-\frac{1}{2} \left(1-\frac{1}{\sqrt{1+4 \Lambda w/s^2}}\right) \right]\\
&=\int_0^{\infty}\frac{y^2}{4 \Lambda (1-2 x)^3} \exp\left(-\frac{1}{4}g_{\Lambda}(x)\, y^2-y\right)  \, dy,
\end{split}
\eeq
where the $w$ integral is performed first, $y=2s$, and
\[ g_{\Lambda}(x) = \frac{x(1-x)}{\Lambda (1-2 x)^2}.\]
The function $g_{\Lambda}(x)$, suggested by perturbation theory, is symmetric
about $x=1/2$, i.e.\ $g_{\Lambda}(x)=g_{\Lambda}(1-x)$.  It includes a scaling
by $\Lambda$, which magnifies the eigenvalue $\lambda_2$, and the value becomes
arbitrarily large whenever the second largest eigenvalue gets close to $1/2$.

This implies the remarkable result that for the
variable $u_2 = g_{\Lambda}(\lambda_2)$ there is a
universal density {\it independent} of $\Lambda$,
\begin{equation}
\begin{split}
\rho_{u_2}(u) &=  \frac{1}{4} \int_0^{\infty} e^{-t} e^{-t^2u/4} t^2 dt\\
&=-\frac{1}{u^2} + \frac{\sqrt{\pi}}{2 u^{5/2}}\left((2 + u)  \exp\left(1\over u \right)  \text{erfc}\left (1\over \sqrt{u} \right)\right).
\end{split}
\label{eq:distrlam2-new-power-law}
\end{equation}

Recall that the complementary error function is
$\text{erfc}(x)=1-(2/\sqrt{\pi}) \int_0^{x} e^{-t^2} dt \approx
1-2x/\sqrt{\pi}$, the approximation being valid for $x \approx 0$.
Thus for $u \gg 1$, the density of $\rho_{u_2}(u)$ has a
power-law tail $\sim \sqrt{\pi}/(2 u^{3/2})$.

\begin{figure}[b]
\includegraphics{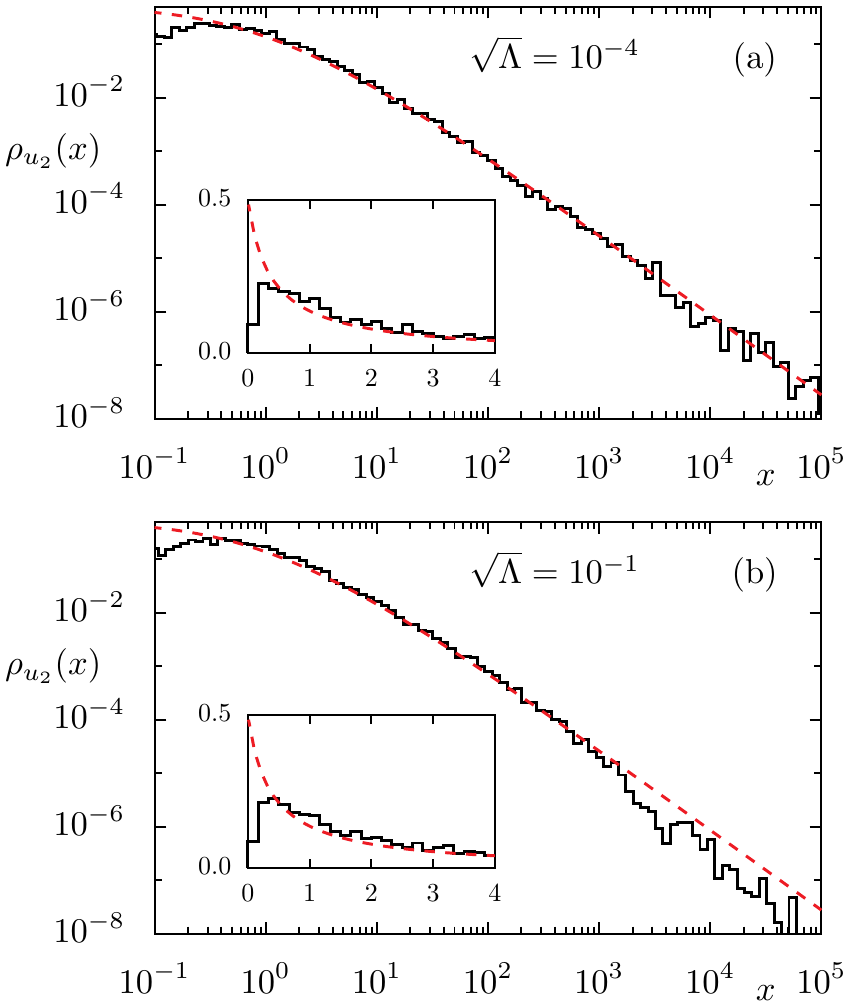}
\caption{\label{fig:distrib-u2-new-power-law}
Probability density $\rho_{u_2}(x)$ for $u_2=g_\Lambda(\lambda_2)$
in a double-logarithmic representation for the coupled kicked rotors.
The dashed curves (red) shows the analytical result
\eqref{eq:distrlam2-new-power-law}.
The insets show the same densities in a linear representation.}
\end{figure}

Figure~\ref{fig:distrib-u2-new-power-law} illustrates the validity of the
power-law tail over several orders of magnitude of $\sqrt{\Lambda}$.
Although the power-law tail remains quite intact even for larger coupling
strengths, such as $\sqrt{\Lambda}=0.1$, deviations are visible around
$u \sim 1$ even for the smallest $\Lambda$ used. This is in the regime when
$\lambda_2 \sim \Lambda$, whereas the average $\lambda_2$ is much higher, being
$\sim \sqrt{\Lambda}$, see Eq.~(\ref{eq:lambda2-avg-approx}).  Thus, it appears
that the current approximations used for the second eigenvalue are not good
enough to capture these very small values accurately.  Indeed, the average is
also calculated to within $\mathcal{O}(\Lambda)$.  The deviations then reflect
the need for higher order perturbation theory.  Note also that
$\overline{u_2 }=\infty$ due to its density having a power-law tail
$u^{-3/2}$, and reflects the fact that the average of $\lambda_2$ is not of
order $\Lambda$.

\subsubsection{Density of the largest eigenvalue}

The largest eigenvalue $\lambda_1$ is related to a sum over many terms each
arising from such heavy--tailed densities and is hence naturally related to
L\'evy stable distributions, through the generalized central limit theorem.
The first equation in Eq.~(\ref{eq:lambda12}) implies
\beq
\label{eq:lam1dist-2a}
1-\lambda_1=\sum_{j=1}^{M}x_j,
\eeq
where each $x_j$ is distributed according to the density
\beq
\label{eq:lam1dist-2b}
\begin{split}
&\rho_{x_j}(x)= \\
&\frac{1}{M}\int_{-M/2}^{M/2} ds \int_0^{\infty}e^{-w} dw \; \delta\left[ x-\frac{1}{2}
\left(1-\frac{1}{\sqrt{1+4 \Lambda w/s^2}}\right) \right]\\
&=\frac{1}{M}\int_{-M/2}^{M/2} e^{-s^2 g_{\Lambda}(x)} \frac{s^2}{\Lambda (1-2 x)^3}\, ds.
\end{split}
\eeq
Note that $s$ is uniformly distributed on $[-M/2,M/2)$ as opposed to being an
exponential as in the case of the second largest eigenvalue, see
Eq.~(\ref{eq:distlamb2-1}).

Observe that
\beq
\label{eq:lam1dist-3a}
g_{\Lambda}(1-\lambda_1)=g_{\Lambda}(\lambda_1) \approx \sum_{j=1}^{M} y_j,\;\; y_j =g(x_j),
\eeq
where the approximation
$g_{\Lambda}(\sum_{j}x_j)\approx \sum_j g_{\Lambda}(x_j)$ is used and justified
as there is no constant term in the Taylor expansion of $g_{\Lambda}$ and the
fact that to leading order $(\sum_{j>1} x_j)^k \approx \sum_{j>1}x_j^k$, where
the cross-terms involving independent random variables have been
neglected. This is similar to the arguments that lead to the approximation in
Eq.~(\ref{eq:ensemble-averaged-moments}).

\begin{figure}[b]
\includegraphics{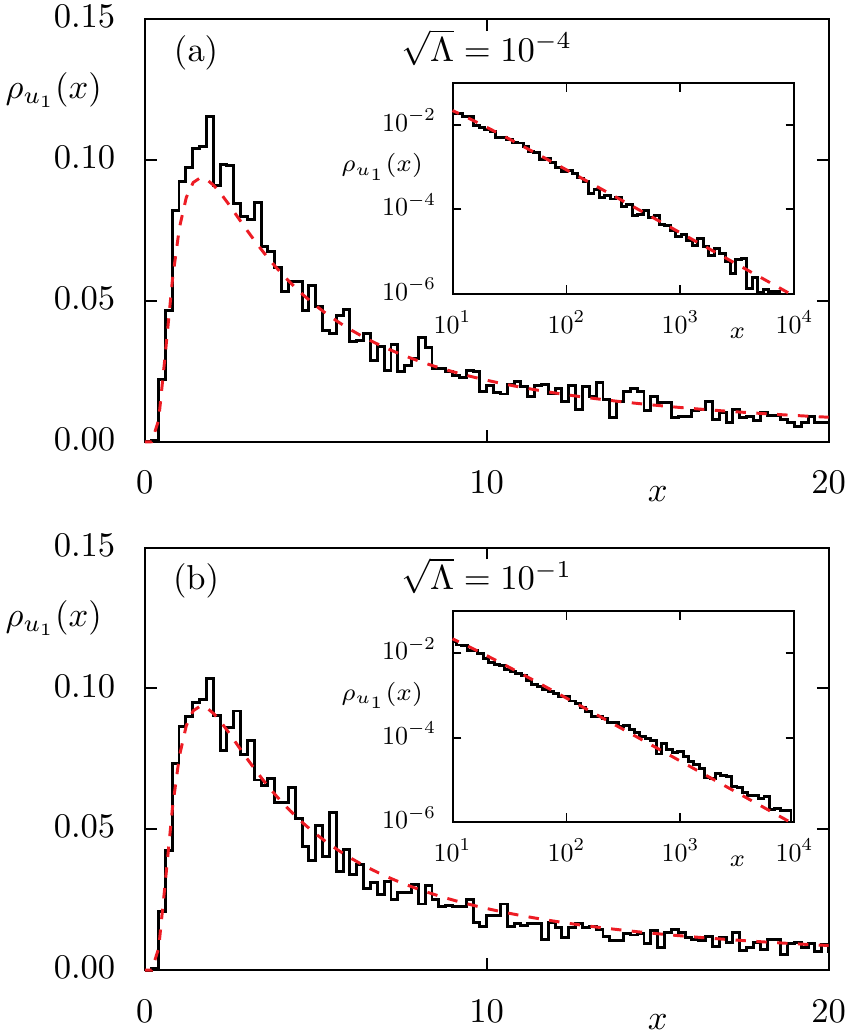}
\caption{\label{fig:distrib-u1-new-levy}
Probability density $\rho_{u_1}(x)$ of $u_1 = g_\Lambda(\lambda_1)$
 for the coupled kicked rotors
in comparison with the L\'evy distribution \eqref{eq:lam1dist-4},
dashed red line.
The insets show the density in a double-logarithmic
representation highlighting the excellent agreement in the tails,
especially for $\sqrt{\Lambda}=10^{-4}$.}
\end{figure}

The density of each $y_j$ is given by
\beq
\label{eq:lam1dist-3b}
\rho_{y_j}(y) = \frac{2}{M} \int_0^{M/2} e^{-s^2 y} \, s^2 \, ds \rightarrow \frac{\sqrt{\pi}}{2M} \frac{1}{y^{3/2}}.
\eeq
If we scale $y_j$ to $\tilde{y}_j=M^2 y_j/\pi$, then the density of
$\tilde{y}_j$ has a tail that is independent of the number of terms summed and
goes as $(1/2) \tilde{y}^{-3/2}$ and $g_{\Lambda}(\lambda_1)$ is distributed as
the sum
\beq
\frac{1}{M^2} \sum_{j=1}^M \tilde{y}_j.
\eeq
Then according to a generalized central limit
theorem~\cite{BouGeo1990,UchZol1999},
the sum limits to the L\'evy distribution
with index $\alpha=1/2$ and scaling constant $\pi/2$. Thus if
$u_1 \equiv g_{\Lambda}(\lambda_1)$, then it is distributed with the L\'evy
probability density
\beq
\label{eq:lam1dist-4}
\rho_{u_1}(u)= \frac{\sqrt{\pi}}{2u^{3/2}} \exp\left(-\frac{\pi^2}{4 u}\right).
\eeq
Shown in Fig.~\ref{fig:distrib-u1-new-levy} is a comparison of this density
with results for the coupled kicked rotors. The power-law tail is again well
reproduced and indicates the large probability with which excursions occur for
the largest eigenvalue away from its unperturbed value.  Therefore, if two
chaotic systems are weakly coupled there is an extended regime in which the
system responds very sensitively with Schmidt eigenvalues being heavy--tailed.
This is reflected in the averages of the eigenvalues deviating greater from
their unperturbed values than expected from a naive perturbation expansion.

\subsubsection{Density of the purity}

The density of the purity $\mu_2$ is a closely related quantity as
\beq
\mu_2= \left(1-\sum_{i>1}\lambda_i\right)^2 + \sum_{i>1} \lambda_i^2 \approx 1-2 \sum_{i>1}\lambda_i (1-\lambda_i) ,
\eeq
where cross-terms (eigenvalue correlations) have been
neglected as before as they are of lower order.
Using Eq.~(\ref{eq:lambda12}) we get
\beq
\frac{1}{2}(1-\mu_2)=\sum_{j=1}^N\left(4+\dfrac{s_j^2}{\Lambda w_j}\right)^{-1} \equiv \sum_{j=1}^N x_j,
\eeq
with the $x_j$ being distributed according to
\beq
\rho_{x_j}(x)=\frac{2}{N} \int_0^{N/2} \exp\left(\frac{-s^2 x}{\Lambda (1-4x)} \right) \frac{s^2}{\Lambda (1-4x)^2} ds.
\eeq
This is therefore similar to the scenario
in Eqs.~\eqref{eq:lam1dist-2a} and \eqref{eq:lam1dist-2b}
except that $g_{\Lambda}(x)$ is now replaced with
\beq
f_{\Lambda}(x)=\frac{x}{\Lambda (1-4x)}.
\eeq
Following the same procedure as for the largest eigenvalue case gives
\beq
\frac{1-\mu_2}{2 \Lambda (2  \mu_2 -1)}=\frac{S_2}{2 \Lambda(1-2 S_2)}
\eeq
and is L\'evy distributed as in Eq.~(\ref{eq:lam1dist-4}). The purity is
written in terms of the linear entropy $S_2$ above which implies that the
entanglement probability density, suitably scaled in the perturbative regime,
is the stable L\'evy distribution.

\begin{figure}[t]
\includegraphics{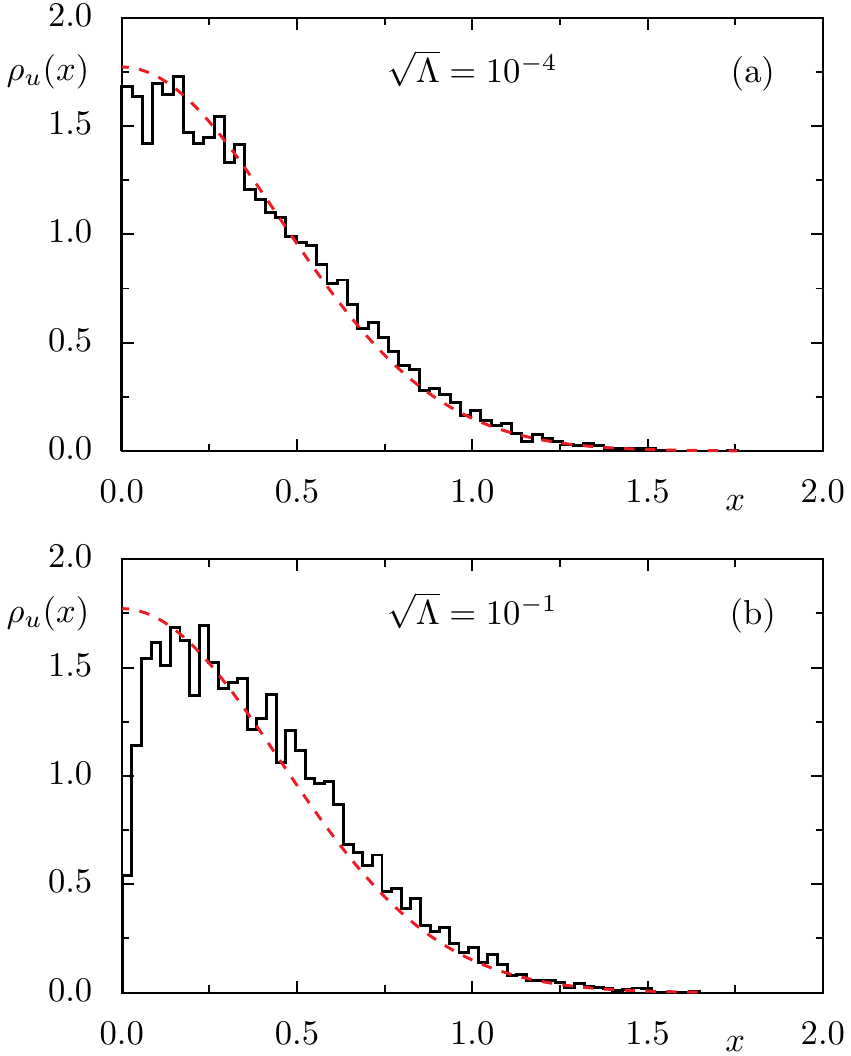}

\caption{\label{fig:distrib-P2}
Probability density $\rho_u(x)$ of $u$ as defined in Eq.~\eqref{eq:u-for-P-2}
based on the purity $\mu_2$ for the coupled kicked rotors
in dependence on $\sqrt{\Lambda}$
in comparison with the half-normal density \eqref{eq:half-normal},
red dashed line.}
\end{figure}

If $y$ is a random variable that is L\'evy distributed as in
Eq.~(\ref{eq:lam1dist-4}), it is easy to see that $u=1/\sqrt{y}$ is distributed
according to the ``half-normal" density given by
\begin{equation} \label{eq:half-normal}
 \rho_u(x)=\sqrt{\pi} \exp(-\pi^2 x^2/4)
 \qquad \text{with $x \ge 0$}.
\end{equation}
Thus displayed in Fig.~\ref{fig:distrib-P2} for comparison with coupled
kicked rotors is the probability density of the related quantity
\beq \label{eq:u-for-P-2}
  u = \sqrt{\frac{2 \Lambda (2 \mu_2-1)}{1-\mu_2}}
    = \sqrt{\frac{2 \Lambda (1- 2S_2)}{S_2}},
\eeq
which should be distributed according to Eq.~\eqref{eq:half-normal}.
Note that these expressions are expected to be valid for $\Lambda \ll 1$,
where $P_2>1/2$. Again one observes that the agreement is very good up to
$\sqrt{\Lambda}=0.01$, while for $\sqrt{\Lambda}=0.1$ clear deviations, in
particular at small $u$, are visible.

\FloatBarrier
\subsection{Non-perturbative regime}

\subsubsection{Density of the purity}

\begin{figure}
\includegraphics{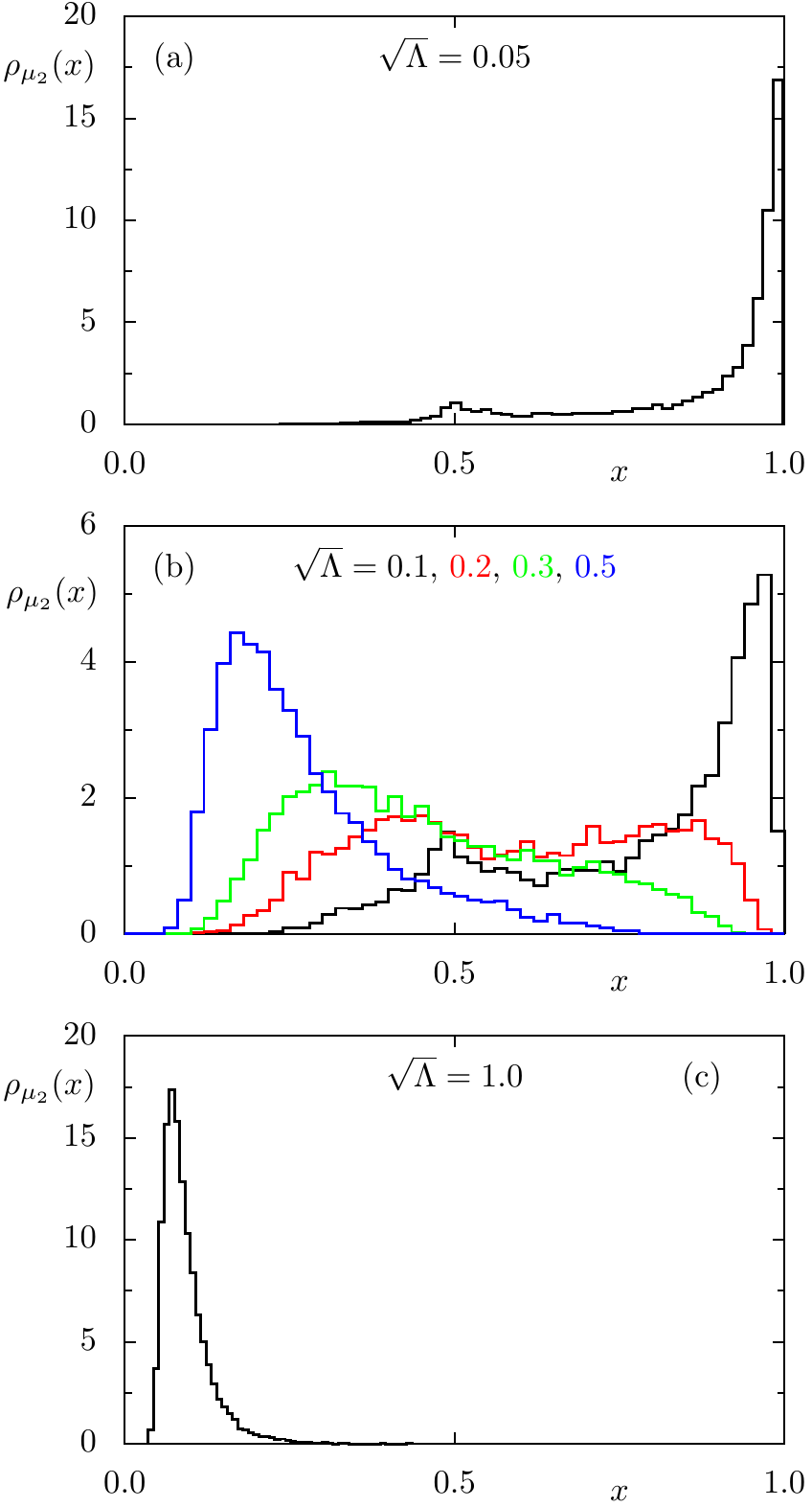}
\caption{\label{fig:distrib-P-alpha}
Probability densities $\rho_{\mu_2}(x)$ of the purity $\mu_2$
for the coupled kicked rotors for
$\sqrt{\Lambda}=0.05, 0.1, 0.2, 0.3, 0.5, 1.0$.}
\end{figure}

Shown in Fig.~\ref{fig:distrib-P-alpha} is the density
$\rho_{\mu_2}(x)$  of the purity itself,
across a wide range of the transition parameter $\Lambda$. It is seen
that at around $\sqrt{\Lambda}=0.1$ a prominent secondary peak appears around
purity $x=1/2$. This corresponds well with the value of the interaction for
which the density of the largest eigenvalue starts to overlap with that of
the second largest in a significant manner as seen earlier in
Fig.~\ref{fig:distrib-lambda-1-2}. For larger values of $\Lambda$, the other
eigenvalues also compete as is illustrated in Fig.~\ref{fig:lambda_i-vs-Lambda}
and decreases the secondary peak's purity further. The densities become
unimodal once again and proceed towards to the random matrix densities
with a mean value around $2/N$.
Although the entire transition remains to be captured, the
next section shows how the mean value and similar mean values for other
entropies can be motivated to evolve in an essentially simple manner.

\subsubsection{Transition of the density of Schmidt eigenvalues}

\begin{figure}
\includegraphics{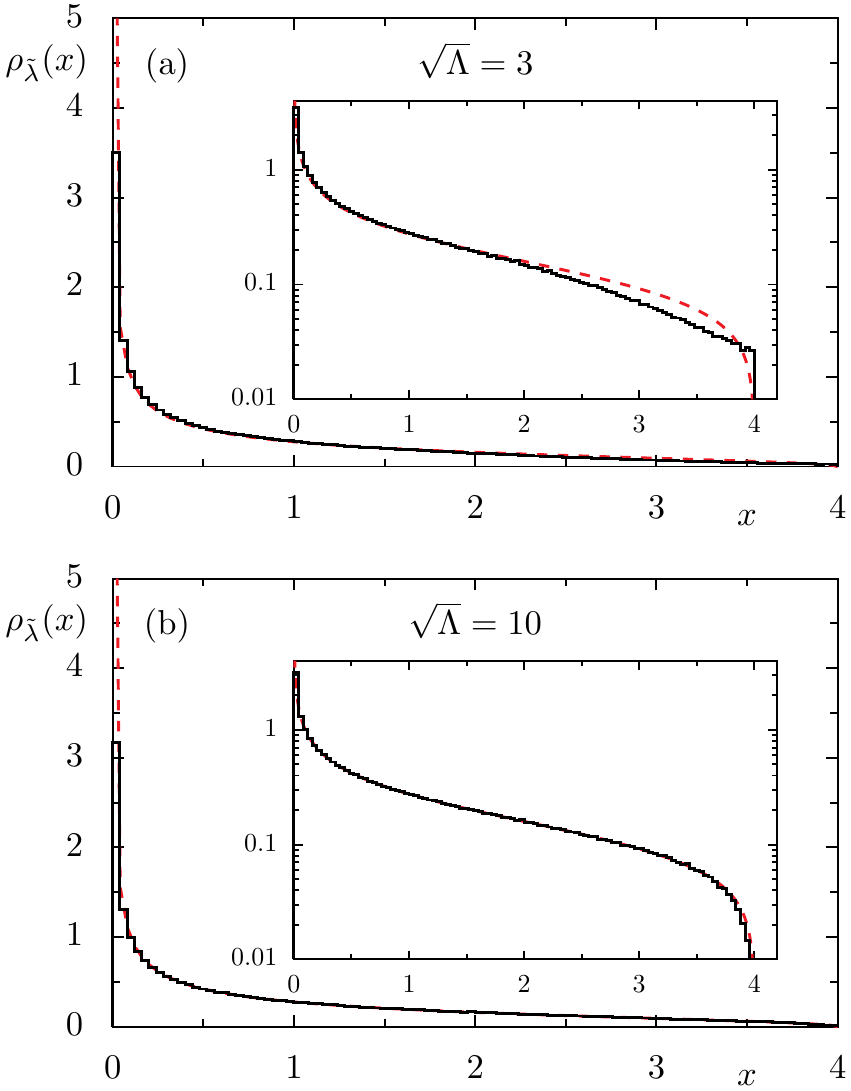}
\caption{\label{fig:distrib-MP}
Density $\rho_{\tilde{\lambda}}(x)$ of all Schmidt eigenvalues,
rescaled according to Eq.~\eqref{eq:tilde-lambda-i-def},
for the coupled kicked rotors (black histogram)
in comparison with the \Marcenko-Pastur distribution
\eqref{eq:marcenko-pastur-law} for $Q=1$, red dashed line.
The inset shows the same data in the semi-logarithmic representation.}
\end{figure}

For the case of strong interaction, i.e.\ large $\Lambda$, one expects that the
statistics of any quantity of interest follows the corresponding
random matrix results.
The approach to this limit can be, as illustrated here, quite distinct.  First,
consider the density of
\begin{equation} \label{eq:tilde-lambda-i-def}
  \tilde{\lambda_i} = N\lambda_i,
\end{equation}
which for $\Lambda$ large enough must follow the \Marcenko-Pastur distribution,
Eq.~\eqref{eq:marcenko-pastur-law}.  Toward this end, consider the density of
the rescaled eigenvalues.  Figure~\ref{fig:distrib-MP} shows the combined
density obtained from $N$ Schmidt eigenvalues for each of the $N^2$ eigenstates
for the coupled kicked rotors.  For $\sqrt{\Lambda} = 3$, deviations are still
clearly visible in the tail of the density visible in the inset.  For
$\sqrt{\Lambda} = 10$ the agreement with the \Marcenko-Pastur distribution is
quite good apart for a small deviation in the tail, whereas for
$\sqrt{\Lambda} = 15$ (not shown) the agreement is excellent.

\begin{figure}
\includegraphics{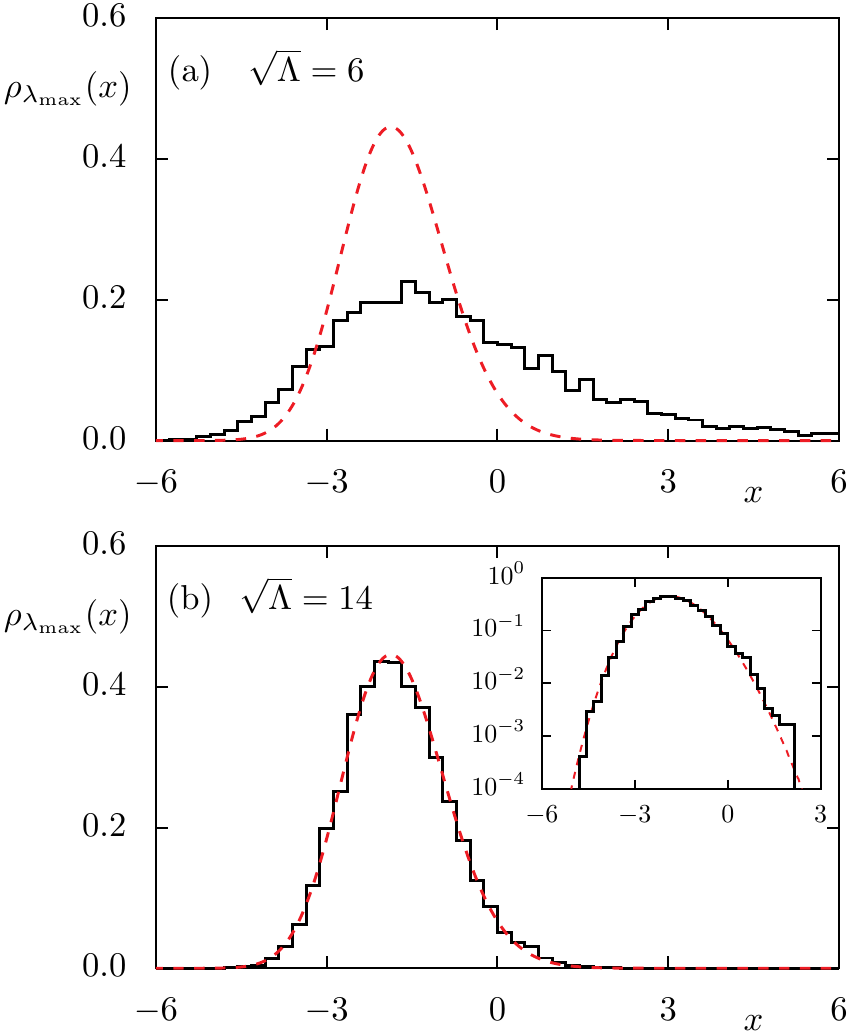}
\caption{\label{fig:lambda-1-distrib-large-coupling}
Density $\rho_{\lambda_{\text{max}}}(x)$ of the largest Schmidt
eigenvalue $\lambda_1$, rescaled according to Eq.~\eqref{eq:tw-rescaling},
for the coupled kicked rotors for different $\sqrt{\Lambda}$.
For comparison the Tracy-Widom distribution, expected in the random matrix theory limit,
is shown (dashed red line).}
\end{figure}

Of particular importance is the probability density of the largest eigenvalue
$\lambda_1$, which has already been considered in the perturbative regime, see
Fig.~\ref{fig:distrib-lambda-1-2}.
In the random matrix theory limit the behavior in the tail
of the \Marcenko-Pastur distribution is governed by $\lambda_1$.
In this limit a Tracy-Widom distribution $F_2$ is expected for the unitary
case~\cite{TraWid1996,EdeRao2005}, if one considers the
appropriately rescaled variable
\begin{equation}
\label{eq:tw-rescaling}
\lambda_{\text{max}}  = \frac{\lambda_1 - 4/N}{2^{4/3} N^{-5/3}},
\end{equation}
see e.g.~\cite[Eq.~(55)]{Nec2007}.
Of interest is how the Tracy-Widom distribution is approached
as the interaction is increased $\sqrt{\Lambda}$.
Figure~\ref{fig:lambda-1-distrib-large-coupling} shows that the approach is
much slower than for any other statistical quantity considered in this paper:
only for about $\sqrt{\Lambda} = 14$ good agreement with the Tracy-Widom
distribution is observed when $N=100$.
Interestingly, we find that for $N=50$ already
for $\sqrt{\Lambda} = 6$ quite good agreement
with the Tracy-Widom distribution is obtained (not shown).
This gives a hint that for this statistics
the transition parameter does not provide the right scaling;
understanding this in detail is left for future investigation.

\begin{figure}
\includegraphics{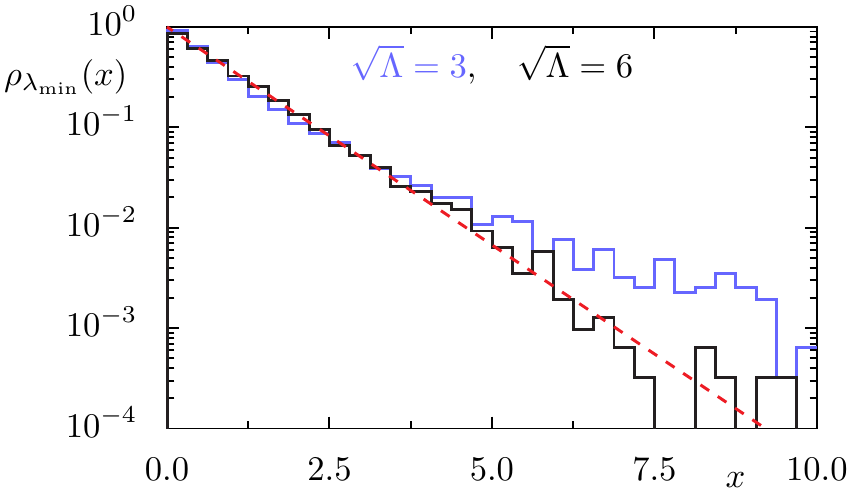}
\caption{\label{fig:distrib_lambda-min}
Density $\rho_{\lambda_{\text{min}}}(x)$ of the
smallest Schmidt eigenvalue $\lambda_N$,
rescaled according to Eq.~\eqref{eq:exp-rescaling},
for the coupled kicked rotors for $\sqrt{\Lambda}=3$ (blue)
and $\sqrt{\Lambda}=6$ (black).
For comparison the exponential density, expected in the random matrix limit,
is shown (dashed red line).}
\end{figure}

For random matrix theory, the density of the
smallest Schmidt eigenvalue $\lambda_N$ is
proven to be exponential~\cite{MajBohLak2008} using the rescaling
\begin{equation} \label{eq:exp-rescaling}
  \lambda_{\text{min}}\equiv N (N^2-1) \lambda_N .
\end{equation}
Figure~\ref{fig:distrib_lambda-min} shows that
for $\sqrt{\Lambda}=3$ the tail of the
density is clearly not following the exponential behavior, but
$\sqrt{\Lambda}=6$ shows good agreement.

\section{Recursively embedded  perturbation theory}
\label{sec:recursive-perturbation-theory}

The analytical expressions obtained thus far are perturbative, i.e.\
$\Lambda \ll 1$, but random matrix theory is expected to reproduce
 the entire transition for
chaotic systems.  The opposite limit, $\Lambda \rightarrow \infty$, is also
likely to allow for analytic calculations.  As the main object of interest has
been the spectra of the reduced density matrices,
the relevant random matrix ensemble is the
fixed-trace Wishart ensemble \cite{ZycSom2001}
for which a variety of results are already known \cite{ZycSom2001,SomZyc2004,KumSamAna2017,MajBohLak2008,MajVer2009,NadMajVer2011}.
Thus it would be interesting to
connect the perturbative regimes with various power laws to
this random matrix regime.
Here, we restrict to the average of moments rather than their probability
densities.

Increasing $\Lambda$ gradually from $0$, the eigenvalues of
the reduced density matrix of an eigenstate of the full system
get added one at a time, see
Fig.~\ref{fig:lambda_i-vs-Lambda} for an illustration in terms
of the averages $\overline{\lambda_i}$.
Therefore this defines successive regimes in which one Schmidt
eigenvalue after another starts to increase significantly away from 0.
In the first regime
there are roughly $\sim \sqrt{\Lambda} N^2$ eigenstates whose reduced density
matrix has two prominent eigenvalues.  Thus for $\sqrt{\Lambda} \ll 1/N^2$ the
largest eigenvalue is still nearly unity for all eigenstates.  For
$\sqrt{\Lambda} \sim 1/N^2$ due to near degeneracies in the system's spectrum,
certain eigenstate pairs suddenly appear in uncorrelated,
widely separated parts of the
spectrum that have two dominant eigenvalues $\lambda_1$ and $\lambda_2$.  They
are responsible for the averages that deviate by an order $\sqrt{\Lambda}$ (as
opposed to $\mathcal{O}(\Lambda)$) from their values in the absence of
interactions.  As $\Lambda$ is continuously increased, clusters of three
levels, whose eigenstates have three significant eigenvalues appear, the third
one being of the order of $\Lambda\ln \Lambda$ in the previous regime of pairs
only.  In this regime, the third Schmidt eigenvalue starts to develop
significance and there is a regime where the fourth is also important and so
on.

This scenario suggests an analysis that can capture the essence of the
successive regimes in the transition.  Consider the first regime and
let the pair of unperturbed states $|kl \kt$ and $|k_1 l_1 \kt$ be
one such doublet creating a pair of eigenstates,
one of whom's Schmidt decomposition is very nearly
\beq
\sqrt{\nu_1} \, |kl \kt+\sqrt{\nu_1'} \, |k_1 l_1 \kt ,
\eeq
where $\nu_1+\nu_1'=1$.  For the other member of the pair, $\nu_{1},\, \nu_1'$
are interchanged, so assume that $\nu_1'\le \nu_1$.  Further increase in the
interaction starts to mix in say $| k_2 l_2 \kt$, such that the Schmidt
decomposition is now approximately
\beq
\sqrt{\nu_2} \left(\sqrt{\nu_1} \,| kl\kt+\sqrt{\nu_1'}\,  | k_1 l_1 \kt \right)+\sqrt{\nu_2'} | k_2 l_2 \kt,
\eeq
again with $\nu_2+\nu_2'=1$ and $\nu_2'\le \nu_2$.  At this stage there are
three prominent Schmidt eigenvalues and the corresponding probabilities:
$\nu_2 \nu_1, \nu_2 \nu_1', \nu_2'$. The process is now iterated, thus
schematically corresponds to a fragmentation process of the interval $[0,1]$
into smaller pieces that add to $1$:
\beq
\begin{split}
 \{\nu_1,\nu_1'\} &\rightarrow \{\nu_2 \nu_1,\nu_2\nu_1',\nu_2'\} \\
& \rightarrow \{\nu_3\nu_2 \nu_1,\nu_3 \nu_2\nu_1',\nu_3\nu_2', \nu_3'\} \rightarrow \cdots,
\end{split}
\eeq
where $\nu_j'+ \nu_j=1$ at all stages, but $\nu_j$ is a random variable. At a
particular generation let the fractions be
$\{\lambda_1, \cdots,\lambda_{K-1}\}$ and the associated moment be
$\mu_{\alpha}=\sum_{j=1}^{K-1}\lambda_j^{\alpha}$.
Then after the next level
\beq \mu'_{\alpha}=\nu_{K}^{\alpha} \mu_{\alpha} +(\nu_{K}')^{\alpha}. \eeq
Thus
\beq
\mu'_{\alpha}-\mu_{\alpha}=-(1-\nu_{K}^{\alpha} -(\nu_{K}')^{\alpha})\, \mu_{\alpha} +(\nu_{K}')^{\alpha}(1-\mu_{\alpha}).
\eeq
Now consider only the leading order of $\overline{\mu_{\alpha}}$ in
Eq.~(\ref{eq:momentsalph}) and assume that $\nu_K$ and $\nu_K'$ have the same
properties as $\lambda_1$ and $\lambda_2$, in particular that their moments
satisfy Eq.~(\ref{highjmoment}) and Eq.~(\ref{eq:ensemble-averaged-moments}).
Note that Eq.~\eqref{highjmoment} is also the moment of $\lambda_2$ as the
other eigenvalues contribute at a smaller order of $\sqrt{\Lambda}$. Putting
these together we get an equation for the averages,
\begin{equation} \label{eq:averages-p-alpha-diff}
  \overline{\mu_{\alpha}' } - \overline{\mu_{\alpha}} = -C(\alpha) \sqrt{\Lambda}\ \overline{\mu_{\alpha}} +{\mathcal O}(\Lambda),
\end{equation}
which suggests that the fragmentation process occurs at a rate proportional to
$\sqrt{\Lambda}$. Taking into account the known behavior of
$\overline{\mu_{\alpha}}$ around $\sqrt{\Lambda}=0$ leads to the differential
equation
\begin{equation} \label{eq:differential-equation}
\frac{\partial \overline{\mu_{\alpha}}}{\partial \sqrt{\Lambda}}
  = -C(\alpha) \overline{\mu_{\alpha}}.
\end{equation}
This recursively embedded perturbative argument suggests a simple exponential
extension of the perturbation expansions.  That is the differential equation
solution
\beq
\label{expcform}
\overline{\mu_{\alpha}} \approx \exp\left(-C(\alpha) \sqrt{\Lambda} \right),
\eeq
is expected to be valid for larger $\Lambda$ than does the linear perturbative
result in Eq.~(\ref{eq:momentsalph}). This supports an exponential decay of
moments, in particular the purity ($\alpha=2$), as the interaction is
increased.

In the  $\Lambda\rightarrow \infty$ limit,
the known random matrix asymptotic (large $N$) result is
\begin{equation}
\overline{\mu_{\alpha}^{\infty}}=\mathcal{C}_{\alpha}/N^{\alpha-1},
\end{equation}
where the $\mathcal{C}_{\alpha}$ are Catalan numbers \cite[\S26.5.]{DLMF},
defined usually for integer value of $\alpha$. The
$\overline{\mu_{\alpha}^{\infty}}$, being the moments of the \Marcenko-Pastur
distribution \eqref{eq:marcenko-pastur-law} are well-defined for all
$\alpha\ge 0$ and hence $\mathcal{C}_{\alpha}$ are well-defined as well.

A simple interpolation between the exponential decay of the moments
and their random matrix value is given by
\beq
\overline{\mu_{\alpha}} \approx \exp\left(-\frac{C(\alpha)}{1-\overline{\mu^{\infty}_{\alpha}}}  \sqrt{\Lambda}\right) \left(1-\overline{\mu^{\infty}_{\alpha}} \right)+\overline{\mu^{\infty}_{\alpha}}.
\label{eq:PalphaRecPert}
\eeq
The arguments of the exponentials in Eq.~\eqref{expcform} get modified by the
asymptotic values so that the small $\Lambda$ perturbative result is unchanged,
yet the asymptotic value is reached.
Using \eqref{eq:Tsallis} this gives the for the entropies
\begin{equation}
\overline{S_{\alpha}(\Lambda)} \approx \left[1-\exp\left(- \frac{C(\alpha)}{ (\alpha-1) \overline{S_{\alpha}^{\infty}}}\sqrt{\Lambda}\right) \right] \overline{S^{\infty}_{\alpha}},
\label{eq:entropy}
\end{equation}
where
\begin{equation}
\overline{S_{\alpha}^{\infty}} =\frac{1-\mathcal{C}_{\alpha} N^{1-\alpha}}{\alpha-1}.
\label{eq:alphaSinf}
\end{equation}
The asymptotic entropies $\overline{S_{\alpha}^{\infty}}$ are reached at the
end of the transition, and although Eq.~(\ref{eq:alphaSinf}) is valid for
$\alpha>1$, it is known that $\overline{S_1^{\infty}}=\ln N-\frac{1}{2}$.  For
$\alpha=1$ one has the important case of the von Neumann entropy,
which, however, has to be treated separately.
Its increase is governed by the limit of
$C(\alpha)/(\alpha-1)$ as $\alpha \rightarrow 1$ which is equal to $\pi^{3/2}$,
consistent with Eq.~(\ref{eq:S-1-perturb}).
Explicitly the von Neumann entropy is
\beq
\overline{S_1(\Lambda)}\approx \left[1-\exp\left(- \frac{\pi^{3/2}}{  \overline{S_{1}^{\infty}}}\sqrt{\Lambda}\right) \right] \overline{S^{\infty}_{1}}.
\label{eq:vN_entropy}
\eeq

\begin{figure}[b]
\includegraphics{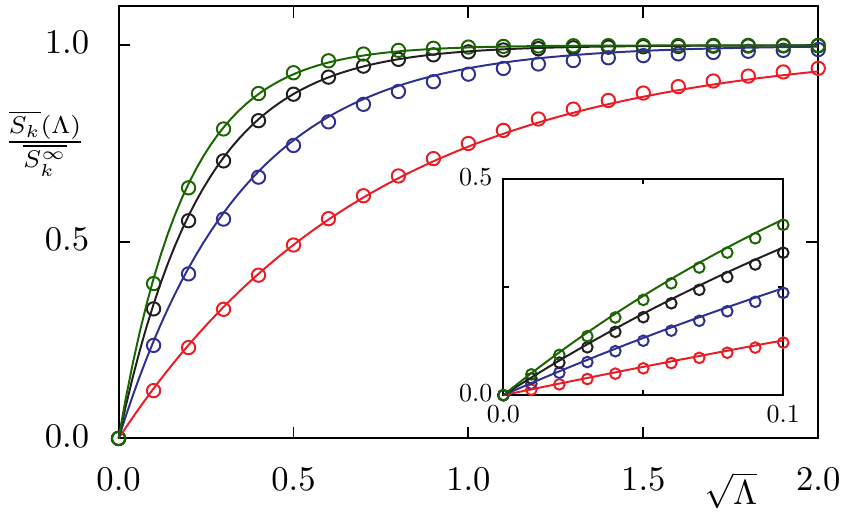}
\caption{\label{fig:entropy}
The average eigenstate entropies $\overline{S_\alpha}$
for von Neumann ($S_1$), and $\alpha=2,3,4$ as a function of $\sqrt{\Lambda}$.
The lowest curve is for von Neumann, and the approach to asymptotic values
is faster for larger $\alpha$ values.  The circles are for the
coupled kicked rotors.  The lines correspond to Eq.~(\ref{eq:entropy}).
The inset is a magnification of the small-$\Lambda$ region.}
\end{figure}

Figure~\ref{fig:entropy} shows the von Neumann entropy and $S_2,S_3,S_4$ for
the coupled kicked rotors.  The agreement is surprisingly good with
Eqs.~\eqref{eq:entropy} and \eqref{eq:vN_entropy}. The inset may be compared
with Fig.~\ref{fig:avg-entropies}, where deviations are visible from the
perturbation theory at values of $\sqrt{\Lambda}$ that is one order of
magnitude smaller.  Overall the exponential continuation along with
the random matrix value seems to give the full transition from uncoupled,
unentangled states to
generic states with nearly maximal entanglement entropy.

\section{Summary and outlook}

The results obtained in this paper apply to a rather general class of weakly
interacting bipartite systems.  The basic assumption is that each of the
subsystems can be considered quantum chaotic in the sense that it can be
described by random matrix statistics for the eigenvalues and eigenvectors.
The random matrix ensemble, Eq.~\eqref{eq:mod}, describes the universal
transition from the statistics of non-interacting systems to those of a fully
interacting case, as long as it is characterized as a function of the unitless
transition parameter $\Lambda$, Eq.~\eqref{eq:transition-parameter}. This
transition is described by a generalized universality class of a dynamical
symmetry breaking nature.  While the subsystems are assumed to follow random
matrix statistics, the interaction between the subsystems is kept general.
Thus, this approach, numerically illustrated for the example of the coupled
kicked rotors, applies equally well to few- and many-body systems, e.g.\
interacting particles in quantum dots, spin chains, coupled quantum maps, and
Floquet systems. Many dynamical symmetry breaking scenarios may be imagined
for many-body systems, e.g., spin-spin interactions or weak environmental
coupling.

A random matrix transition ensemble is used to derive the entanglement
properties of two interacting subsystems. Based on a perturbative treatment,
the single universal transition parameter $\Lambda$ is determined in terms of
the interaction.  Starting from the non-interacting case, $\Lambda=0$, the
entanglement between the subsystems increases to being nearly maximal, i.e.\ to
that of random states in the full Hilbert space.  Quantitatively the
entanglement may be characterized by the purity, HCT entropies, and the von
Neumann entropy.  All these can be computed using the Schmidt eigenvalues
$\lambda_i$ of the reduced density matrix.  Based on perturbation theory
explicit expressions for the largest and second largest eigenvalue of the
reduced matrix are obtained.  Using an appropriate regularization, predictions
for the averages $\overline{\lambda_1}$ and $\overline{\lambda_2}$ are derived,
which are valid for small values of $\sqrt{\Lambda}$.  In this regime good
agreement with numerical results for the coupled kicked rotors is found.
Furthermore, the average moments $\sum_{j>1} \overline{\lambda_j^\alpha}$ and
$\overline{\lambda_1^\alpha}$ are obtained perturbatively up to order
$\mathcal{O}(\Lambda)$.  Based on these moments, a perturbative prediction for
the entanglement entropies, Eq.~\eqref{eq:momentsdefn}, is obtained. A
comparison with numerical results for the coupled kicked rotors shows that the
perturbative results provide a good approximation up to about
$\sqrt{\Lambda}=0.3$. This indicates that the theoretical results are
robust, in the sense that deviations from ideal behavior, such as discussed
in App.~\ref{app:distribution-of-w} do not harm the agreement.
A particularly interesting moment is $\overline{\mu_{1/2}}$ as this
indicates the distance of the eigenfunction to the closest maximally entangled
state.  Here the perturbative result gives good agreement with numerics up to
$\sqrt{\Lambda}=0.6$.

Going beyond the average behavior of the
eigenvalue moments and entanglement entropies
the probability densities of the eigenvalues of the reduced
density matrices are studied.
The probability densities of $\lambda_1$ and $\lambda_2$,
in dependence on the transition parameter $\Lambda$, show substantial tails.
In the perturbative regime, a suitable rescaling of $\lambda_2$
has a universal density which is independent of $\Lambda$,
and shows a power-law tail, found over several orders of magnitude.
For the density of a rescaled $\lambda_1$,
which is a sum of heavy-tailed densities,
a generalized central limit theorem implies the L\'evy distribution,
which also shows a power-law tail.
Closely related is the density of the linear entropy $S_2$,
which also follows the L\'evy distribution.
In turn the purity $\mu_2=1-S_2$, again rescaled, shows a half
normal density, which is also well seen in the numerical
results for the coupled kicked rotors with good agreement
at small $\sqrt{\Lambda}$ and deviations showing up for
$\sqrt{\Lambda}=10^{-1}$.

In the non-perturbative regime the density of the purity
shows a transition from a small amount of entanglement, $\mu_2$ near 1,
to large entanglement, $\mu_2$ around $2/N$.
The density of rescaled Schmidt eigenvalues
approaches the \Marcenko-Pastur distribution.
Of particular interest is the density of the largest
eigenvalue $\lambda_1$, which in the limit of large interaction
follows the Tracy-Widom distribution.
Interestingly, the approach to this limit is much
slower than for any of the other statistical quantities considered
in this paper, with good agreement with numerics
only found for $\sqrt{\Lambda}=14$.
In contrast, the approach
of the density of the smallest Schmidt eigenvalue
to the exponential density of the random matrix theory
limit appears to be faster.

To obtain a description of the behavior of the entanglement entropies for
the whole transition, a recursively embedded
perturbation theory is invoked.
This leads to a remarkably simple expression, Eq.~\eqref{eq:PalphaRecPert},
which actually captures the essence of the entire
transition. It is found to be in very good agreement
with the numerical results for the coupled kicked rotors,
see Fig.~\ref{fig:entropy}.

Moreover, eigenstate localization as measured
by the inverse participation ratio
is connected to the linear entropy after subsystem averaging,
as was shown in \cite{LakSriKetBaeTom2016}.
Thus using a spectral averaging, this can be used as sensitive
detector of non-ergodic behaviors.
An interesting future direction
concerns the relation between entropies and eigenvector moments.

It is interesting, that the rate of growth of entropy $S_{\alpha}$
as a function of the transition parameter,  has precisely the
same form as the $\alpha^{\text{th}}$ multifractal exponent
of eigenstates of critical random matrix ensembles describing
the physics of Anderson transitions~\cite{MirEve2000,EveMir2008}.
For example compare the leading order term in
Eq.~(\ref{eq:hct1}) with Eq.~(3.20) of \cite{EveMir2008}.
This is no coincidence as both are derived as a consequence of
a degenerate perturbation theory that starts from Poissonian spectra.
However, beyond this similarity, although there are power-laws in the
systems we are investigating, if there is criticality in some sense remains
to be seen. The kinds of random matrix ensembles used in this context
thus far are single-particle
ones ~\cite{EveMir2008,BogGir2011a,BogGir2011}
and it is interesting to see if criticality can arise as a
consequence of coupling chaotic systems.

An important future application of the results for
the random matrix transition ensemble
is that this provides a means to detect non-universal
behaviors in terms of deviations from the obtained universal results.
For example for the situation in which one or both of the subsystems
shows a mixed phase space or some kind of localization of eigenstates,
the results given here represent
an important limit to compare with. It is reasonable
to expect that at any given interaction strength the entropies derived herein
are an upper bound in typical systems such as in the coupled kicked rotors.

The application of the results in the context of weakly interacting many-body
system, such as spin-chains, will be of particular
interest. This work raises the question how to determine
the transition parameter $\Lambda$ in such situations
and then investigate the universality of statistics in dependence
on $\Lambda$.
Besides investigating the properties of spectra
and eigenfunctions, also the time-evolution of
wave-packets is expected to follow a universal behavior
in dependence on the transition parameter.
This should also pave the way for experimental studies,
e.g.\ using cold-atom experiments.
Such results will be of importance for
dynamical systems, quantum information, and condensed matter theory.

\begin{acknowledgments}

We would like to thank Roland Ketzmerick for useful discussions.
AL thanks  Ferdinand Evers for a discussion on critical ensembles.

\end{acknowledgments}

\begin{appendix}

\section{Perturbation theory with unitary matrices}
\label{app:perturbation}

Here we derive the perturbation theory results for eigenvalues
and eigenfunctions of unitary matrices.
The eigenvalue results of this appendix can be found in the book
by Peres~\cite{Per2002}.  Consider a unitary operator of the form
\beq
\label{eq:Uabappendix}
\begin{split}
 \cU(\eps) & = (U_A \otimes U_B)\, U_{AB}(\eps) \\
U_{AB}(\eps) & =\exp (i\epsilon V),
\end{split}
\eeq
where ${\cal H}_A,{\cal H}_B$ (dimensions $N_A\le N_B$, respectively) are
assumed to be symmetry reduced subspaces, and hence there are no systematic
degeneracies.  It has unperturbed eigenstates separately for the individual
subspaces of the direct product space given by,
\beq
U_A|j^A\rangle = e^{i\theta^A_j}|j^A\rangle \qquad U_B|k^B\rangle
               = e^{i\theta^B_k}|k^B\rangle
\eeq
and therefore
\beq
(U_A \otimes U_B)|j^A\rangle|k^B\rangle
        = e^{i\theta_{jk}}|j^A\rangle|k^B\rangle \qquad
\theta_{jk}  = \theta^A_j+\theta^B_k
\eeq
The eigenstates of the full system can be denoted
by a double index label $jk$.  Then
\beq
\label{uabeigen}
(U_A \otimes U_B)\, U_{AB}(\eps) |\Phi_{jk}\rangle
   = e^{i\varphi_{jk}}|\Phi_{jk}\rangle
\eeq
where in the $\epsilon\rightarrow 0$ limit,
$|\Phi_{jk}\rangle \rightarrow |j^A\rangle|k^B\rangle\ ( = |jk\rangle)$ and
$\varphi_{jk}\rightarrow \theta_{jk}$.  The standard perturbation approach is
to insist that $\langle jk|\Phi_{jk}\rangle =1$, thereby implying that
$\langle \Phi_{jk}|\Phi_{jk}\rangle> 1$, and it is necessary to renormalize the
eigenvectors in a second step.  The ansatzes for the unnormalized eigenvector
and eigenangle are
\beq
\label{anzatses}
\begin{split}
|\Phi_{jk}\rangle
   & = |jk\rangle +\sum_{n=1}^\infty \epsilon^n |\Phi_{jk}^{(n)}\rangle \\
\varphi_{jk}
   & = \sum_{n=0}^\infty \epsilon^n \varphi_{jk}^{(n)},
\end{split}
\eeq
where it remains to calculate the $n^{th}$-order corrections
$|\Phi_{jk}^{(n)}\rangle$, and the eigenangle, $\varphi_{jk}^{(n)}$.  The
equations to all orders can be generated by projecting a complete set of states
using the unperturbed basis onto Eq.~\eqref{uabeigen},
\beq
\begin{split}
& \sum_{j^\prime k^\prime=1}^{N_AN_B} |j^\prime k^\prime\rangle \langle j^\prime k^\prime |(U_A \otimes U_B)\, U_{AB}(\eps) |\Phi_{jk}\rangle  = \\
& e^{i\varphi_{jk}} \sum_{j^\prime k^\prime=1}^{N_AN_B} |j^\prime k^\prime\rangle \langle j^\prime k^\prime | \Phi_{jk}\rangle = \\
& e^{i\theta_{j^\prime k^\prime}} \sum_{j^\prime k^\prime=1}^{N_AN_B} |j^\prime k^\prime\rangle \langle j^\prime k^\prime|U_{AB}(\eps)|\Phi_{jk}\rangle \ ,
\end{split}
\eeq
then substituting the anzatses of Eq.~\eqref{anzatses}, the exponential form of
the interaction in Eq.~\eqref{eq:Uabappendix}, and creating a hierarchy of
equations by collecting terms of equal order in $\epsilon$.  Each equation must
hold individually for any particular term in the summation $jk^\prime$ above.
The first equation is,
\beq
\begin{split}
&e^{i\theta_{jk}} \left[ i \delta_{jk,j^\prime k^\prime}  \varphi^{(1)}_{jk} +  \left( 1- \delta_{jk,j^\prime k^\prime} \right) \langle j^\prime k^\prime|\Phi_{jk}^{(1)}\rangle \right] \\
& = e^{i\theta_{{j^\prime k}^\prime}} \left[ \left( 1- \delta_{jk,j^\prime k^\prime} \right) \langle j^\prime k^\prime | \Phi_{jk}^{(1)}\rangle +  i\langle j^\prime k^\prime | V | jk \rangle \right]
\end{split}
\eeq
The $jk=j^\prime k^\prime$ term gives the expected relation
\beq
\varphi^1_{jk} = \langle jk | V | jk\rangle
\eeq
and the $jk \ne j^\prime k^\prime$ terms give
\beq
\langle j^\prime k^\prime|\Phi_{jk}^{(1)}\rangle
   = \frac{i\langle j^\prime k^\prime | V | jk \rangle}
          {e^{i\left(\theta_{jk} - \theta_{j^\prime k^\prime} \right)} -1} .
\eeq
In the limit of $N_A \rightarrow \infty$, only differentially small angle
differences are perturbative and this equation reduces to the familiar form
arising with perturbation theory of Hamiltonian systems,
\beq
\langle j^\prime k^\prime |\Phi_{jk}^{(1)}\rangle
   = \frac{\langle j^\prime k^\prime | V | jk \rangle}
          {\theta_{jk} - \theta_{j^\prime k^\prime} },
\eeq
where the eigenangles appear instead of the eigenenergies.

The equation for the second order corrections is longer and it is simpler to
isolate the $\delta_{jk,j^\prime k^\prime}$ terms from those with
$1- \delta_{jk,j^\prime k^\prime}$.  First, the $\delta_{jk,j^\prime k^\prime}$
terms lead to
\beq
i \varphi_{jk}^{(2)} - \frac{(\varphi_{jk}^{(1)})^2}{2} = i \langle jk | V | \Phi_{jk}^{(1)}\rangle - \frac{1}{2}\langle jk | V^2 | jk\rangle.
\eeq
Substituting in the solutions from the first order equations gives
\beq
\varphi_{jk}^{(2)} = \sum_{j^\prime k^\prime\ne jk}^{N_AN_B}\left[ \frac{i}{e^{i\left(\theta_{jk}-\theta_{j^\prime k^\prime} \right)}-1}+\frac{i}{2} \right] \left|\langle j^\prime k^\prime | V | jk \rangle \right|^2 .
\eeq
Perhaps it is less than immediately obvious that the second
order eigenangle correction is real, but consider that
\beq
\begin{split}
\frac{i}{e^{i\left(\theta_{jk}-\theta_{j^\prime k^\prime}\right)}-1}+\frac{i}{2} &= \frac{\sin\left(\theta_{jk}-\theta_{j^\prime k^\prime}\right)}{4\sin^2\left(\frac{\theta_{jk}-\theta_{j^\prime k^\prime}}{2}\right)} \\
& =\frac{1}{2}\cot \left(\frac{\theta_{jk}-\theta_{j^\prime k^\prime}}{2}\right)
\end{split}
\eeq
and also in the $N_A\rightarrow \infty$ limit,
the usual form of the second order eigenvalue correction is recovered
\beq
\varphi_{jk}^{(2)} = \sum_{j^\prime k^\prime\ne jk}^{N_AN_B} \frac{\left|\langle j^\prime k^\prime | V | jk\rangle \right|^2}{\theta_{jk}-\theta_{j^\prime k^\prime}} .
\eeq
That leaves the equation resulting from the second order terms
without the $j^\prime k^\prime=jk$ terms.
After a bit of algebra and grouping terms, this gives
\beq
\begin{split}
& \langle j^\prime k^\prime |\Phi_{jk}^{(2)}\rangle = \frac{i}{e^{i\left(\theta_{jk}-\theta_{j^\prime k^\prime}\right)}-1} \times \\
& \left[ \sum_{j^{\prime\prime} k^{\prime\prime}\ne jk}^{N_AN_B} \frac{\langle j^\prime k^\prime | V | j^{\prime\prime} k^{\prime\prime} \rangle\langle j^{\prime\prime} k^{\prime\prime} | V | jk \rangle \sin\left(\theta_{jk}-\theta_{j^{\prime\prime} k^{\prime\prime}}\right)}{4\sin^2\left(\frac{\theta_{jk}-\theta_{j^{\prime\prime} k^{\prime\prime}}}{2}\right)} +  \right. \\
& \qquad \qquad \left. \frac{\langle j^\prime k^\prime | V | jk \rangle\langle jk | V | jk \rangle \sin\left(\theta_{jk}-\theta_{j^\prime k^{\prime}}\right)}{4\sin^2\left(\frac{\theta_{jk}-\theta_{j^\prime k^\prime}}{2}\right)} \right],
\end{split}
\eeq
which also reduces to the expected expressions in the $N_A\rightarrow \infty$
limit.  Renormalizing the eigenfunction alters the unit coefficient in front of
$|jk\rangle$ in the expansion to
\beq
|\Phi_{jk}\rangle = \left[ 1-\frac{\epsilon^2}{8} \sum_{j^\prime k^\prime \ne jk}^{N_AN_B}\frac{\left|\langle j^\prime k^\prime | V | jk\rangle \right|^2}{\sin^2\left(\frac{\theta_{jk}-\theta_{j^\prime k^\prime}}{2} \right)}\right] |jk\rangle + ...
\eeq
also with the expected $N_A\rightarrow \infty$ limit.

\section{Transition parameter}
\label{app:universal-scaling-parameter}

In this appendix we derive expression \eqref{eq:lambda1-new}
for the transition parameter $\Lambda$.
Consider the bipartite unitary operator in Eq.~\eqref{eq:Uab}
for which we want to calculate $v^2$,
the mean square off-diagonal matrix element
of $U_{AB}(\epsilon)$ in the basis in which $U_A$ and $U_B$ are diagonal.
To simplify notation we omit the dependence on $\epsilon$.
Moreover, for convenience we assume that
$U_{A,B}$ are sampled independently from the CUE (of $N_A \times
N_A$, and $N_B \times N_B$ unitary matrices respectively) and
averaging is also performed over this ensemble.

In the unperturbed basis, denote the off-diagonal matrix element
\[ z_{kl;k'l'}=\sum_{m=1}^{N_A} \sum_{n=1}^{N_B} u_{km}
u_{k'm}^{*} w_{ln} w_{l'n}^{*}
(U_{AB})_{mn},\] where $(k,l)\neq (k',l')$, and $(U_{AB})_{mn}=\langle
mn|U_{AB}|mn\rangle$ is a diagonal matrix element due to the
interaction and $u, w$ are two independent $N_A, N_B$--dimensional
unitary matrices, respectively, which diagonalize the unperturbed part.
The averaging is done
over the Haar measure on unitary matrices.
Using the independence of $u$ and $w$, and the relation
\[
  \overline{u_{km}u_{k'm'}u^*_{k'm}u^*_{km'}}
  = \frac{\delta_{mm'-1/N_a}}{N_A^2 -1},
\]
gives
\begin{equation}
\begin{split}
  ~ &  v^2 = \overline{|z_{kl;k'l'}|^2} \\
  ~ &=
\frac{\sum_{m,m'}^{N_A}
\sum_{n,n'}^{N_B} \left[
\delta_{mm'}\delta_{nn'} -
\frac{\delta_{nn'}}{N_A}-\frac{\delta_{mm'}}{
N_B} +
\frac{1}{N_AN_B}\right]
 }{(N_A^2-1)(N_B^2-1)}\\
  & ~\quad \times (U_{AB})_{mn}(U_{AB})^*_{m'n'}.
\end{split}
\end{equation}
This series can be rewritten as
\begin{equation}
\begin{split}
& v^2=\dfrac{N_A^2 N_B^2}{(N_A^2-1)(N_B^2-1)} \times \\
& \left(1+ \left|\frac{\tr U_{AB}}{N_A N_B}
  \right|^2  - \frac{1}{N_A}\, \left\|\frac{U^{(A)}}{N_B}\right\|^2
     -\frac{1}{N_B}\, \left\|\frac{U^{(B)}}{N_A}\right\|^2 \right),
\end{split}
\end{equation}
where $U^{(A)} = \tr_{B} U_{AB}$, $U^{(B)} = \tr_{A} U_{AB}$,
and $\| X\| = \tr (X X^\dagger)$ is Hilbert-Schmidt norm.

The transition parameter $\Lambda = v^2/D^2$,
where $D=\frac{2\pi}{N_A N_B}$ is the mean level spacing,
then takes the form
\begin{equation}
\begin{split}
& \Lambda=\dfrac{N_A^3 N_B^3}{4 \pi^2(N_A^2-1)(N_B^2-1)} \times \\
& \left(1+ \left|\frac{\tr U_{AB}}{N_A N_B}
  \right|^2  - \frac{1}{N_A}\, \left\|\frac{U^{(A)}}{N_B}\right\|^2
     -\frac{1}{N_B}\, \left\|\frac{U^{(B)}}{N_A}\right\|^2 \right).
\end{split}
\label{eq:lambda11}
\end{equation}

Next we evaluate $\Lambda$ explicitly for the model in
Eq.~\eqref{eq:Uab} by performing the averaging over  $\xi_j$.
For this we treat every term in the sum in Eq.~\eqref{eq:lambda11} separately,
\begin{equation*}
\begin{aligned}
 \overline{\left| \frac{\tr U_{AB}}{N_AN_B}\right|^2}
&=\frac{1}{N_A^2N_B^2} \overline{\sum_{j,j'=1}^{N_AN_B} e^{2\pi i
\epsilon \xi_{j}}  e^{-2\pi i \epsilon
\xi_{j'}}} \\
  &=\frac{1}{N_A^2N_B^2} \left[
\sum_{j=1}^{N_AN_B} 1 + \sum_{j \neq j} \overline{e^{2\pi i \epsilon
(\xi_{j}-\xi_{j'})}}\right] .
\end{aligned}
\end{equation*}
As the $\xi_j$'s are independent uniform random variables in $[-1/2,1/2)$,
their difference $z=\xi_j-\xi_{j'}$ has the probability density function,
\begin{equation}
 \rho_{z}(x) = \begin{cases} 1 + x  &\text{ for } -1< x <0\\
         1 - x &\text{ for } 0\leq x <1\\
         0 & \text{ otherwise}.
        \end{cases}
\end{equation}
Using
$\int_{-1}^1 e^{2\pi i \epsilon x} \rho_z(x)dx
=\frac{\sin^2 (\pi \epsilon)}{\pi^2 \epsilon^2}$
gives
\begin{equation*}
 \overline{\left| \frac{\tr U_{AB}}{N_AN_B}\right|^2}
=\frac{1}{N_A N_B} + \left( 1 - \frac{1}{N_AN_B}\right) \frac{\sin^2
(\pi \epsilon)}{\pi^2 \epsilon^2} .
\end{equation*}
Similarly,
\begin{equation}
 \begin{aligned}
  \overline{\frac{1}{N_A}\, \left\|\frac{U^{(A)}}{N_B}\right\|^2} &=
\frac{1}{N_B} + \left( 1 - \frac{1}{N_B}\right) \frac{\sin^2
(\pi \epsilon)}{\pi^2 \epsilon^2}\\
\overline{\frac{1}{N_B}\, \left\|\frac{U^{(B)}}{N_A}\right\|^2}
&=\frac{1}{N_A} + \left( 1 - \frac{1}{N_A}\right) \frac{\sin^2
(\pi \epsilon)}{\pi^2 \epsilon^2}.
 \end{aligned}
\end{equation}
Inserting these results in Eq.~\eqref{eq:lambda11}
finally gives the result
\begin{equation}
\Lambda=\dfrac{N_A^2 N_B^2}{4 \pi^2(N_A + 1)(N_B + 1)}
 \left[1-\dfrac{\sin^2 (\pi \epsilon)}{\pi^2\epsilon^2} \right],
\end{equation}
as stated in Eq.~\eqref{eq:lambdarmt}.

\section{Probability density of matrix elements}
\label{app:distribution-of-w}

\begin{figure}[h]
\includegraphics[width=8.6cm]{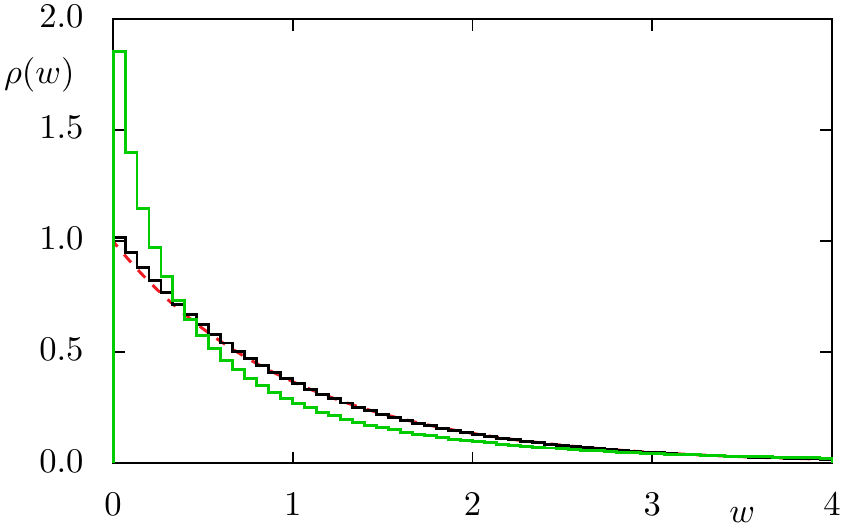}
\caption{\label{fig:off-diag-mat-KR}
Probability density $\rho(w)$ of the (normalized) off-diagonal matrix
elements for the random matrix transition ensemble and coupled kicked rotors
for $N=100$.
For the random matrix ensemble the (black) histogram  matches the
exponential quite well, Eq.~\eqref{eq:off-diagonal-matrix-elements}
(red dashed line), as expected.
The density for the coupled kicked rotors (green histogram)
deviates quite a bit.}
\end{figure}

Consider the density of the off-diagonal matrix elements
$|\br jk | V | j^\prime k^\prime \kt|^2$ for the random matrix ensembles and
for the coupled kicked rotors.  The set of vectors $\{| j k \kt\}$ is the
eigenbasis of the two uncoupled systems and $V$ the symmetry breaking
interactions in Eqs.~\eqref{vjj},\eqref{eq:intpot}, respectively.  It is taken
for granted that for the ensemble, Eq.~\eqref{eq:mod}, after rescaling by the
mean absolute square, $v^2$, there is a unit mean random variable
$w = |\br jk|V |j^\prime k^\prime \kt|^2/v^2$, which follows an exponential
probability density, Eq.~\eqref{eq:off-diagonal-matrix-elements}.  The quality
of this assertion is demonstrated in Fig.~\ref{fig:off-diag-mat-KR}, which
shows that it is very good.

The situation is more complicated for the coupled kicked rotors ($N=100$),
which after all is an actual dynamical system.
The histogram for the kicked rotors
deviates quite a bit from the expected exponential behavior.  This could be a
reflection of a correlation between the matrix elements and eigenvectors,
imperfect ergodicity of the system, or the lack of true randomness of the
interaction, all of which would generally happen,
at least to some extent for a real dynamical system,
as opposed to a member of a random matrix ensemble.
Nevertheless, for all the derived universal results throughout the paper, the
coupled kicked rotors followed the theory quite well.  It has long been known
within random matrix theory that many results are robust in the sense that
deformed ensembles or ones with non-Gaussian matrix elements, or many other
kinds of alterations still lead to the same fluctuation
properties.  It appears that even though the coupled kicked rotors do not
approximate a member of the ensemble perfectly, they lead to the same
results for the quantities studied in this paper,
and thus lie within the range of robustness of the theory.

\end{appendix}

\vfill\eject

\bibliographystyle{cpg_unsrt_title_for_phys_rev}

\bibliography{abbrevs,extracted,references}

\end{document}